\documentclass[final,3p,times,twocolumn]{elsarticle}
\usepackage{enumerate}
\usepackage[cmex10]{amsmath}
\usepackage[font=footnotesize]{subfig}
\usepackage{fixltx2e}
\usepackage{multirow} 
\usepackage{booktabs}

\usepackage{amssymb} 
\usepackage{gensymb} 
\usepackage{algorithm}
\usepackage{algpseudocode} 
\usepackage{comment} 
\usepackage{url}
\usepackage{flushend}
\usepackage{epstopdf}
\usepackage{epsfig}
\usepackage{xcolor}

\journal{Computer Networks}

\begin{document}

\begin{frontmatter}

%
\title{Smartphone-based geolocation of Internet hosts}


\author[label1]{Gloria Ciavarrini}
\ead{gloria.ciavarrini@for.unipi.it}
\author[label2]{Valerio Luconi}
\ead{valerio.luconi@iit.cnr.it}
\author[label1]{Alessio Vecchio\corref{cor1}}
\ead{alessio.vecchio@unipi.it}

\address[label1]{Dip. di Ingegneria dell'Informazione\\ 
Universit\`a di Pisa\\
Largo L. Lazzarino 1, 56122 Pisa, Italy} 

\address[label2]{Istituto di Informatica e Telematica\\
Consiglio Nazionale delle Ricerche\\
Via G. Moruzzi 1, 56124 Pisa, Italy}

\cortext[cor1]{Corresponding author.}

\begin{abstract}

The location of Internet hosts is frequently used in distributed applications
and networking services. Examples include customized advertising, distribution
of content, and position-based security. Unfortunately the relationship between
an IP address and its position is in general very weak.  This motivates the
study of measurement-based IP geolocation techniques, where the position of the
target host is actively estimated using the delays between a number of
landmarks and the target itself.  This paper discusses an IP geolocation method
based on crowdsourcing where the smartphones of users operate as landmarks.
Since smartphones rely on wireless connections, a specific delay-distance model
was derived to capture the characteristics of this novel operating scenario.

\end{abstract}


\end{frontmatter}

\section{Introduction}

Many distributed applications and networking services may benefit from knowing
the geographical position of Internet hosts.  For instance, with such
information, content can be customized depending on the user's position, or a
geographically close replica can be selected when downloading large amounts of
data. Alternatively, the position of a host can be used to restrict on-line
transactions to trusted areas, or to determine the source of cybercrimes.  The
IP address of a host, unfortunately, provides little information about its
position~\cite{Freedman:2005:GLI:1251086.1251099}. A number of databases map IP
addresses with their believed coordinates. Examples include GeoIP by
MaxMind~\cite{maxmind}, IP2Location~\cite{ip2location}, and
IPInfoDB~\cite{ipinfodb}. Some IP geolocation databases are filled using
administrative information (e.g. extracted from Whois entries). The approach
based on administrative information suffers from two main problems: $i)$
information is added by hand by network administrators, thus it may be out of
date; $ii)$ location information is sometimes provided at the organization
level; thus, in case of large administrative domains, it may not be accurate.
Several studies show that the use of administrative information may lead to
errors in the order of several thousand kilometers~\cite{laki10:model,
laki11:spotter, pad01:investigation, Poese:2011:IGD:1971162.1971171}. 

Active IP geolocation methods estimate the position of an Internet host by
performing network measurements. Usually, the observed network parameter
is the delay between the host to be localized (the target) and a number of
hosts with known location (called landmarks). Some methods convert delays into
distances and then use geometrical techniques, such as multilateration, to
compute the position of the target on a global reference system. Other methods
rely on the concept of similarity in the network distance space, and the target
is co-located with the host with known position that exhibits the most similar
delay pattern.

Landmarks collect delay measurements towards the target by sending probes and
acquiring timestamps. They are generally coordinated by a server, which is also
in charge of running a centralized localization algorithm. Almost all the
approaches devised so far use the hosts of academic/research platforms as
landmarks. PlanetLab is the most used platform, as it allows researchers to run
distributed experiments on a wide scale~\cite{planetlab}. In addition, the
position of hosts participating in the PlanetLab network is known.

This paper discusses a smartphone-based IP geolocation method that operates
according to crowdsourcing principles
~\cite{howe2008crowdsourcing,Choffnes:2010:CSN:1851275.1851228}: smartphones
provided by common users are enrolled as measuring devices and used as
landmarks. 
The contribution of this work with respect to existing IP geolocation
techniques is threefold. First, the use of mobile devices as landmarks, which
is made possible by their GPS units, represents an unexplored possibility in
the field of IP geolocation. Second, a delay-distance model that takes into
account the presence of wireless access links is presented; this widens
previous knowledge where wired-only links were considered.  Third, the
experimental platform is composed of crowdsourced devices not belonging to
research facilities; this makes experimental results (and related discussion)
closer to real world operating conditions. 

The paper is organized as follows: Section~\ref{lab:relwork} provides some
background about active IP geolocation; Section~\ref{lab:loc} presents
the main elements of the localization method; in Section~\ref{lab:calibration},
the adopted delay-distance model is discussed; in Section~\ref{sec:results}
experimental results are shown; Section~\ref{lab:conclusions} concludes the
paper.

\section{Related work}
\label{lab:relwork}

Due to their practical relevance, IP geolocation methods received significant
attention during the last years.  A number of methods are based on
administrative information and static registries. Notable examples of
information sources include the Domain Name System (DNS) and Whois.  RFC 1876
standardizes the format of DNS LOC records for experimental
purposes~\cite{RFC1876}. By using these records network administrators can
specify latitude, longitude, and altitude of network resources.  The well known
Whois directory service maps IP addresses to the organizations they belong
to~\cite{RFC3912}.  The physical address (in terms of country, city, street) of
organizations is also stored in Whois databases and it is used to derive
location information.  These databases suffer from two main problems.  First,
they are filled with human-generated information and, as a consequence, they
may contain obsolete entries (as reported in~\cite{zhang2006dns}). For
instance, there is no incentive for network administrators in updating a DNS
LOC entry when an IP address is re-used on a machine with a different location.
Second, an organization may be responsible for a large number of IP addresses
spread over a wide geographic area; in such circumstances the use of Whois may
introduce significant approximations, as a large block of IP addresses is
associated to a single administrative office.

A variation of these techniques is adopted by GeoTrack, which uses DNS-based
information to infer the position of the target
host~\cite{pad01:investigation}.  The addresses of router interfaces along a
network path are converted in names via DNS. Such names frequently contain
city, country or airport codes, which are used to infer the location of the
target.  Unfortunately, DNS names of routers do not follow standardized rules
and this reduces the possibility of applying this technique in different
scenarios.

Active IP geolocation methods rely on end-to-end delay measurements for
determining the position of the target. In particular, landmark hosts send
probes towards the target to collect delay measurements. In some cases, delays
are converted into physical distances; distances are then combined according to
a geometric technique for estimating the position of the target. In other
cases, delays are directly used to locate the target according to network
similarity metrics.

\subsection{Methods based on geometric techniques}

Constraint-based Geolocation (CBG) uses an approach derived from
multilateration \cite{Gueye:2006:CGI:1217687.1217693}. Each landmark estimates
the distance from the target by measuring the round trip time (RTT).  The
minimum RTT is converted into distance using a linear model calibrated for each
landmark (by measuring the delays towards all other landmarks). The estimated
distance is assumed to be an upper bound with respect to the real distance,
because errors in delay measurements are always additive. In practice, each
landmark defines a circular region; the intersection of circular regions
defines a convex region where the target is supposed to be. The position of the
target is calculated as the center of mass of the intersection. 

Spotter is an IP geolocation system based on a probabilistic approach that does
not require landmark-specific calibration~\cite{laki11:spotter}. Spotter relies
on a delay-distance model where the distribution of distances for a given delay
is independent from the position of landmarks. This improves resiliency to
measurement errors and anomalies. Spotter uses PlanetLab nodes as landmarks. 

In~\cite{Dong201285}, the relationship between distance and delay is modeled
according to a segmented polynomial. Authors collected RTT data using traceroute,
with PlanetLab nodes acting as both landmarks and targets. Calibration of each
landmark is carried out dividing data in five regions using a clustering
technique. Then the coefficients of polynomials are computed.  Estimated
distances are used to calculate the position of targets using semidefinite
programming, an approach borrowed from localization in wireless sensor
networks. Some weighting schemes are compared: no weight (all landmarks have
the same importance when localizing a target), inversely proportional to
distance (close landmarks are more important than distant ones), sum weighted
(the ratio between the distance of a given landmark with respect to the sum of
distances of all landmarks is considered).  Results confirm that the polynomial
model provides better distance estimation with respect to the linear one.
Moreover, the inversely proportional weight scheme improves localization
accuracy.

Topology-based Geolocation is an IP geolocation method that improves
delay-based techniques leveraging network
topology~\cite{Katz-Bassett:2006:TIG:1177080.1177090}. The topology of the
network is explored using traceroute, which enables collection of information
about intermediate routers. Other tools are used to cluster network interfaces
(in case they belong to the same router). Delays are used to constrain the
position of hosts: hard constraints are generated using the maximum speed of
light in fiber, soft constraints are generated using the delay associated to
Internet paths and segments.

IP geolocation techniques have also been adapted to perform geolocation of
data: in this case the goal is to localize the host and, at the same time,
obtain proofs of data possession. The system presented
in~\cite{Gondree:2013:GDC:2435349.2435353} can be used to understand if a cloud
provider relocates data to a remote data center, possibly violating legal or
commercial agreements.  Geolocation is based on CBG, using per-landmark
calibration. Delay measurements are not carried out using ICMP probes (or,
broadly speaking, small packets). On the contrary, relatively large
data blocks (up to 32KB) are used to ensure data possession. An experimental evaluation 
was carried out both on a PlanetLab-based scenario and a cloud-based one.
Results are similar to the ones obtained by the original CBG. Authors observed
that the selection of landmarks has enormous importance on localization accuracy.

\subsection{Methods based on network similarity}

GeoPing localizes the target host by selecting the nearest neighbor in delay
space. The delay vector registered by a number of landmarks is compared to
existing data to find the best match (in terms of Euclidean distance), then the
position of the nearest neighbor is returned as the target's
position~\cite{pad01:investigation}. An experimental evaluation was 
carried out in North America, using 14 probing machines and 256 passive
landmarks.

In~\cite{5235373} a statistical geolocation approach is discussed. A profile of
each landmark is obtained by measuring its delay towards the set of remaining
landmarks. Then the joint probability density function of delay and distance is
calculated using a kernel density estimator. A force-directed method is used as
an approximate algorithm to maximize the likelihood of the target location
given the delay measurement data. An experimental evaluation of the proposed
approach was carried out using 85 nodes of the PlanetLab network in the
continental U.S.A. Results show a better localization accuracy with respect to
CBG.

In~\cite{Ziviani2005503} the accuracy of a geolocation method based on delay
similarity is analyzed when varying the placement of passive landmarks and
probing machines. The authors evaluate different strategies: random, geographic
(based on agglomerations), and demographic (based on user concentrations).
Different methods for measuring similarity are studied and compared. In
particular, the authors discuss three methods: one based on distance, one based
on cosine similarity, and one based on correlation.  Manhattan distance is the
one that provides the best performance. The dataset used for the evaluation was
collected by means of the RIPE infrastructure (55 hosts, one-way delay
measurements). The geolocalization procedure places the target host where the
nearest landmark is located.

\subsection{Discussion and motivation}

Almost all IP geolocation studies rely on measurements collected in homogeneous
environments (research/academic networks and testbeds), where delay values are
relatively stable and reliable. Moreover, in all cases, only wired links have
been considered. 

We propose a novel approach, not explored in existing literature, where common
smartphones are used as landmarks. We believe this study can be a significant
addition to the landscape of IP geolocation methods, especially considering the
always growing role of smartphones as general computing and communication
devices. The smartphones used for IP geolocation do not belong to a research
facility or a single organization, instead they are provided by volunteers who
join the system according to crowdsourcing principles. These devices belong to
autonomous systems\footnote{An autonomous system is a group of networks with a single
routing policy, run, in most cases, by a single network
operator~\cite{Hawkinson96:rfc1930}.} at the fringes of the Internet, thus our scenario is more
representative, in terms of operating conditions, of a real-world geolocation
system.  Since measurements are collected via wireless links, a specific
calibration of the delay-distance model has been carried out.  In particular,
not only we derived a new delay-distance model that is specific for paths with
wireless access, but we also devised the possibility of using the type of
access (Wi-Fi, 3G, 4G) to improve the delay-distance relationship.  Both these
aspects are not covered by existing literature, and are sufficiently general to
be used also in other contexts, possibly not related to IP geolocation.

\section{Smartphone-based IP geolocation} 
\label{lab:loc}

\begin{figure}[t!] 
\centering
\includegraphics[width=0.8\columnwidth]{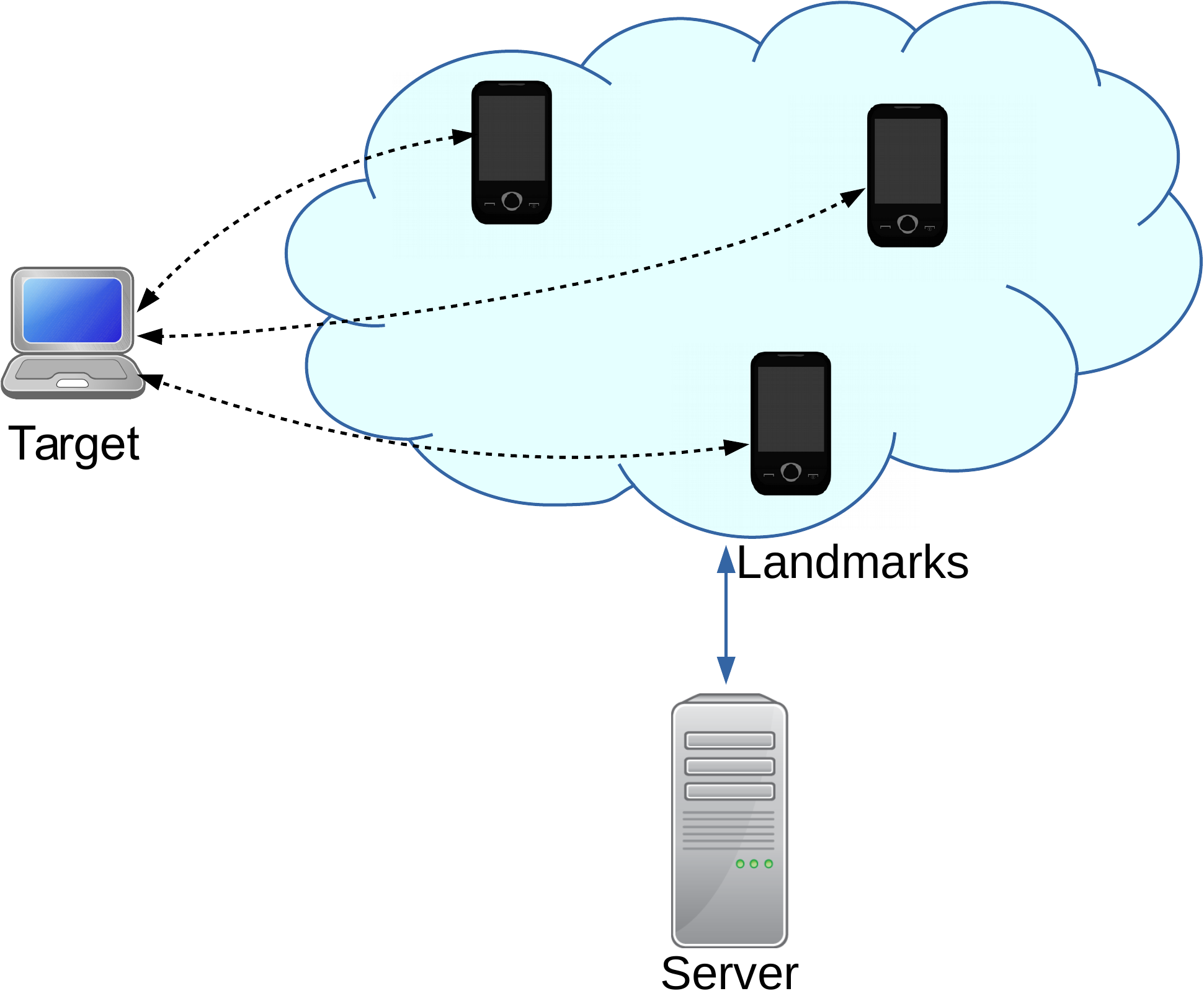} 
\caption{System architecture. A central server remotely triggers smartphones to collect measurements towards the target.}
\label{fig:architettura} 
\end{figure}

The use of smartphones as landmarks in IP geolocation is motivated by their
ability to self-localize using the GPS unit. Moreover, their number and
distribution is continuously increasing and this makes them an appealing
platform for the development of geographically distributed applications.  In
the proposed method, participation to the IP geolocation activities takes place
on a volunteer basis, according to a crowdsourcing-based paradigm. Devices remain
under control of their respective owners, but they can be remotely triggered by
a central server to carry out the requested delay measurements
(Figure~\ref{fig:architettura}). The set of devices involved in localizing a
given host is not constant, as it depends on their availability at the time of
measurements. 

\begin{figure}[t!] 
\begin{algorithmic} 
\State $\mathcal{S} \gets C_{1}$ 
\ForAll {$i \in \{2, \ldots, N\}$} 
  \State $\mathcal{I}_i \gets \mathcal{S} \cap C_{i}$
  \If {$\mathcal{I}_i \neq \emptyset$} 
    \State $\mathcal{S} \gets \mathcal{I}_i$
  \EndIf 
\EndFor 
\end{algorithmic}
\caption{Algorithm used to compute $\mathcal{S}$, the region where the target is supposed to be.} 
\label{alg:localize}
\end{figure}

\subsection{Method}

Let $\mathcal{L} = \{L_{1}, L_{2}, ..., L_{N}\}$ be the set of $N$ landmarks
participating in measurements towards the target. Let $\mathcal{M}_{i} =
\{m_{i,1}, m_{i,2}, \ldots, m_{i, K}\}$ be the set of RTTs between the $i$-th
landmark ($L_{i}$) and the target, where $K$ is the number of RTTs collected by
landmarks. The end-to-end delay between two hosts can be roughly decomposed in
the following factors: transmission delay, processing delay, queuing delay, and
propagation delay. To reduce the ``noise'' introduced by queuing and
processing, each landmark selects the minimum value $\hat{m_{i}}$ of collected
RTTs ($\hat m_{i} = min (\mathcal{M}_{i})$).  

The minimum RTT is then used to estimate the distance $r_i$ between
$L_{i}$ and the target: 
\begin{equation} 
  \label{eq:radius} r_i = f(\hat{m_i})
\end{equation} 
where $f()$ is a function that models the delay-distance relationship. Function
$f()$  is calculated as discussed in Section~\ref{lab:calibration}.

Note that in an environment that includes wireless links, delays are
characterized by increased variability with respect to wired-only networks
(e.g. because of collisions and higher probability of transmission errors).
The problem of collecting accurate delays is exacerbated by the use of
smartphones as measuring elements. In fact, hardware resources and mobile
operating systems are not very suitable for collecting high-resolution
timestamps; reasons include the presence of energy saving mechanisms, the
execution of code only in user space, and the presence of other networked
applications on the device. These inaccuracies in collecting delays produce
errors in estimated distances. 

Distances $\{r_1, ..., r_N\}$ are then combined according to an algorithm
derived from CBG~\cite{Gueye:2006:CGI:1217687.1217693}.  More precisely, in the
proposed method, a circular region with radius $r_i$ is centered at the
position of landmark $L_i$. Let us call $C_{i}$ this region.  Let us also
define $\mathcal{C} = \{C_{1}, C_{2}, \ldots, C_{N}\}$ as the list of all
circular regions ordered according to their radius.

\begin{figure}[t!] 
\centering
\includegraphics[width=0.6\columnwidth]{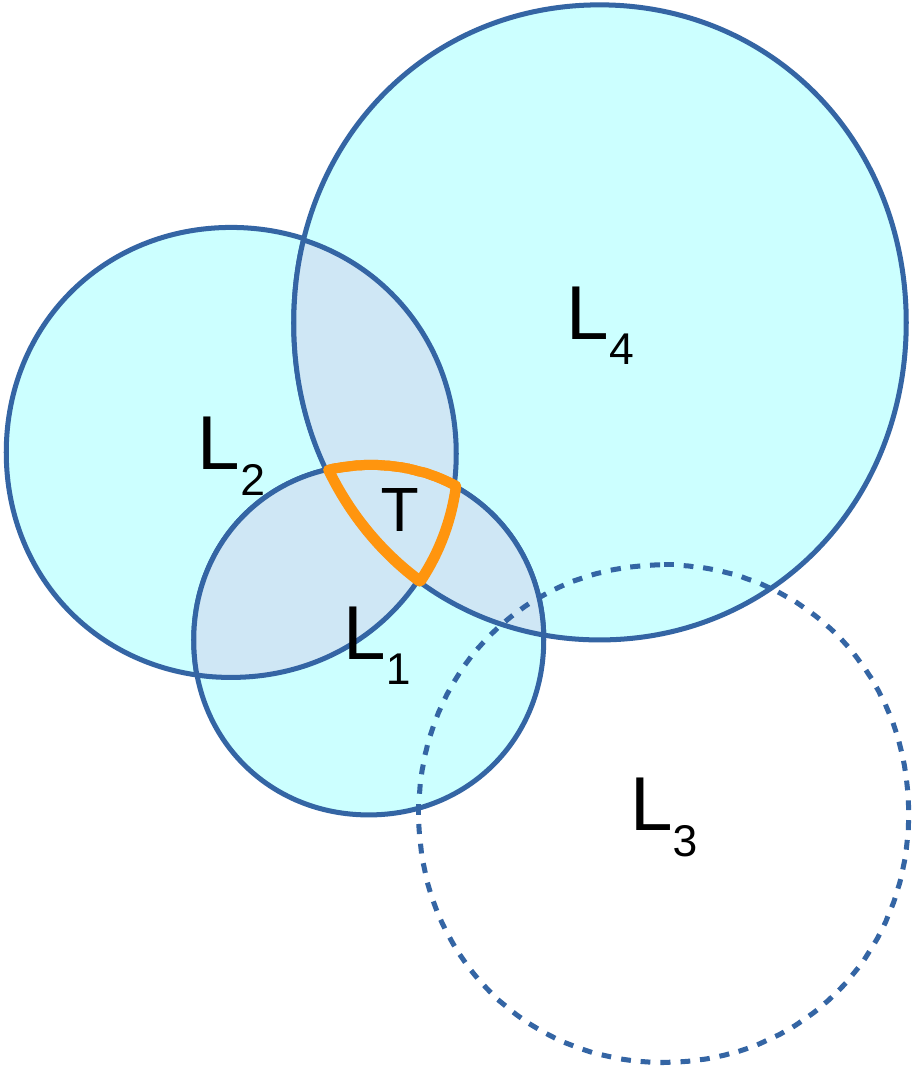} 
\caption{An example of localization with four landmarks; because of distance underestimation $L_3$ does not contribute to localize the target (T).}
\label{fig:algorithm-example} 
\end{figure}

The region $\mathcal{S}$ where the target is supposed to be is computed by
processing the elements of $\mathcal{C}$ according to their order, i.e. from
the smallest to the largest. In particular, $\mathcal{S}$ is initialized with
$C_{1}$ as it is the circular region with smallest radius.  Subsequently, the
intersection $\mathcal{I}_i$ between $C_{i}$ and $\mathcal{S}$ is computed
($\mathcal{I}_i \leftarrow C_{i} \cap \mathcal{S}$).  If $\mathcal{I}_i$ is
null then $C_{i}$ is discarded: some measurements are affected by large errors
and hence not all constraints can be simultaneously satisfied (the algorithm
prefers to discard larger circular regions with respect to smaller ones).  If
$\mathcal{I}_i$ is not null, then $\mathcal{S} \leftarrow \mathcal{I}_i$: the
new value of $\mathcal{S}$ includes the constraints imposed by $C_{i}$ and the
previous ones.  This process is repeated for all elements in $\mathcal{C}$.
The pseudo-code of the algorithm used for computing $\mathcal{S}$ is reported
in Figure~\ref{alg:localize}.  
After having calculated $\mathcal{S}$, its barycenter is used to estimate the
position of the target.

Figure~\ref{fig:algorithm-example} illustrates an example of localization using
four landmarks ($L_1$, ..., $L_4$), ordered with increasing radius.  The
intersection $\mathcal{I}_2$ between $\mathcal{S}$ and $C_{2}$ is not null,
thus this region is used to restrict $\mathcal{S}$ from its initial value
(which was equal to $C_{1}$).  $C_{3}$ has no intersection with
$\mathcal{I}_2$ thus it is not used to further restrict $\mathcal{S}$.
Finally, the intersection between $C_{4}$ and the previously determined region
is used as the final value of $\mathcal{S}$.  The barycenter of $\mathcal{S}$
provides the estimated position of the target.

The proposed localization method is based on CBG because the latter is rather
simple, provides good performance, and was adopted as a reference algorithm in
a number of other works. As mentioned, CBG executes a calibration phase for
every landmark: RTTs are collected towards all other landmarks and the
coefficient used for delay-distance conversion is chosen as the smallest value
that does not generate under-estimations.

Our method differs from the original CBG for the following aspects:

\begin{itemize}

\item There is no per-landmark calibration. Since landmarks are mobile devices,
they are connected to the Internet from possibly different autonomous
systems and locations. Thus, per-device
calibration does not make much sense, as the operating conditions of the device
during calibration can be largely different from the ones at runtime. Instead,
we explored the possibility of performing delay-distance calibration depending
on the wireless technologies used for accessing the network (i.e.  Wi-Fi, UMTS,
HSPA, etc.). This form of calibration is probably less accurate than the one
\`a la CBG, as it coalesces information produced by all landmarks that use the
same technology. However, at the same time, it is less complex, as there is no
need to calibrate new landmarks when they join the system (in a
crowdsourcing-based scenario, where devices join and leave the system
dynamically, calibrating every new device is not feasible). 

\item Circular regions are ordered and used one by one. In particular, regions
are ordered according to their radius; this because larger distances are generally
affected by larger errors.  The algorithm starts using the smallest
region, as it is the one that is probably affected by the smallest error, then
considers the subsequent ones and tries to compute the intersection
incrementally. 

\end{itemize}

It is worth to note that average distances are generally used in combination
with multilateration based approaches. Nevertheless, we found that in the
considered scenario, which is characterized by possibly significant distance
estimation errors, multilateration accuracy can be drastically reduced by the
presence of few largely erroneous measurements. Conversely, maximum distances
are used in approaches based on computing an admissible region where the target
must be located. The latter solution cannot be easily applied in an operational
environment characterized by increased heterogeneity: being delay measurements
affected by a relatively large dispersion, the geometrical constraints that can
be generated are overly loose. The localization procedure here proposed tries
to solve the IP geolocation problem by combining elements from both approaches.
Constraints are softer with respect to other systems, as they may be violated
because they are generated from mean distances. If a constraint cannot be
satisfied, it is simply discarded. For the same reason constraints are ordered
according to the radius of circular regions. Small delay values are
characterized by reduced errors if compared to large values.  Thus the use of
ordered soft constraints tries to favor measurements characterized by good
accuracy with respect to coarse ones.

\subsection{Collection of delay data}

Smartphones involved in collecting network measurements have been enrolled
using the Portolan platform, a crowdsourcing-based system aimed at monitoring
and studying very large-scale
networks~\cite{gregori14:smartphone,gregori13:sensing}.  Smartphones that
participate in Portolan activities are remotely instructed about the
measurement tasks they have to carry out.  Results are then sent to a central
server where they are saved for later processing and analysis. Portolan
provides a number of tools that can be used for characterizing the network
under observation in terms of topology, bandwidth, and delay.  To be part of
the Portolan platform, volunteers just have to install a standard app on their
devices. Portolan is available for a number of operating systems (Android,
Linux, Mac OS X, Windows), but for this study only Android-based devices have
been used.

\begin{figure*}[t] \centering 
\subfloat[Landmarks]{
\includegraphics[width=0.9\columnwidth]{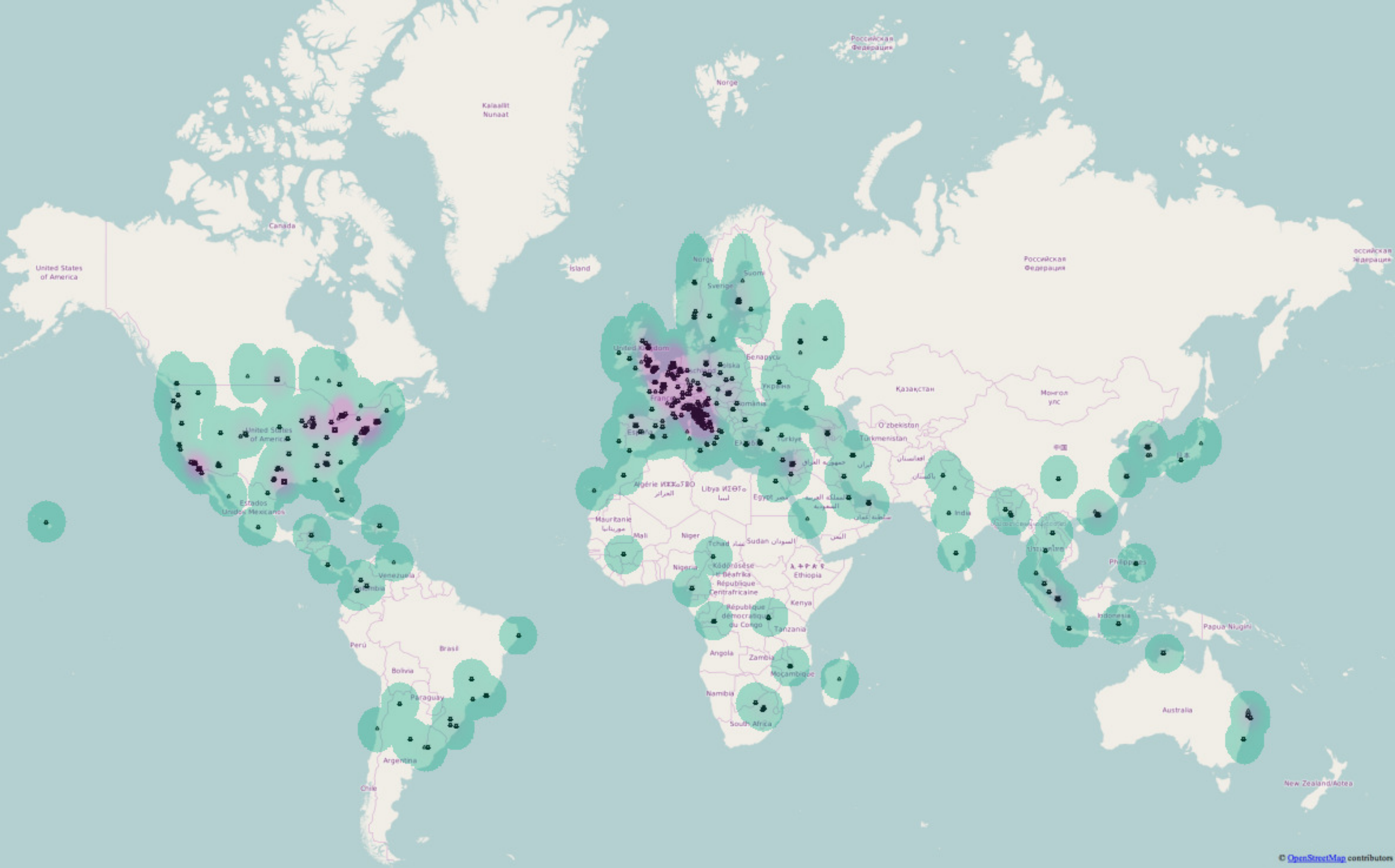}
\label{fig:world-landmarks} } 
\hspace{1cm}
\subfloat[Targets]{
\includegraphics[width=0.9\columnwidth]{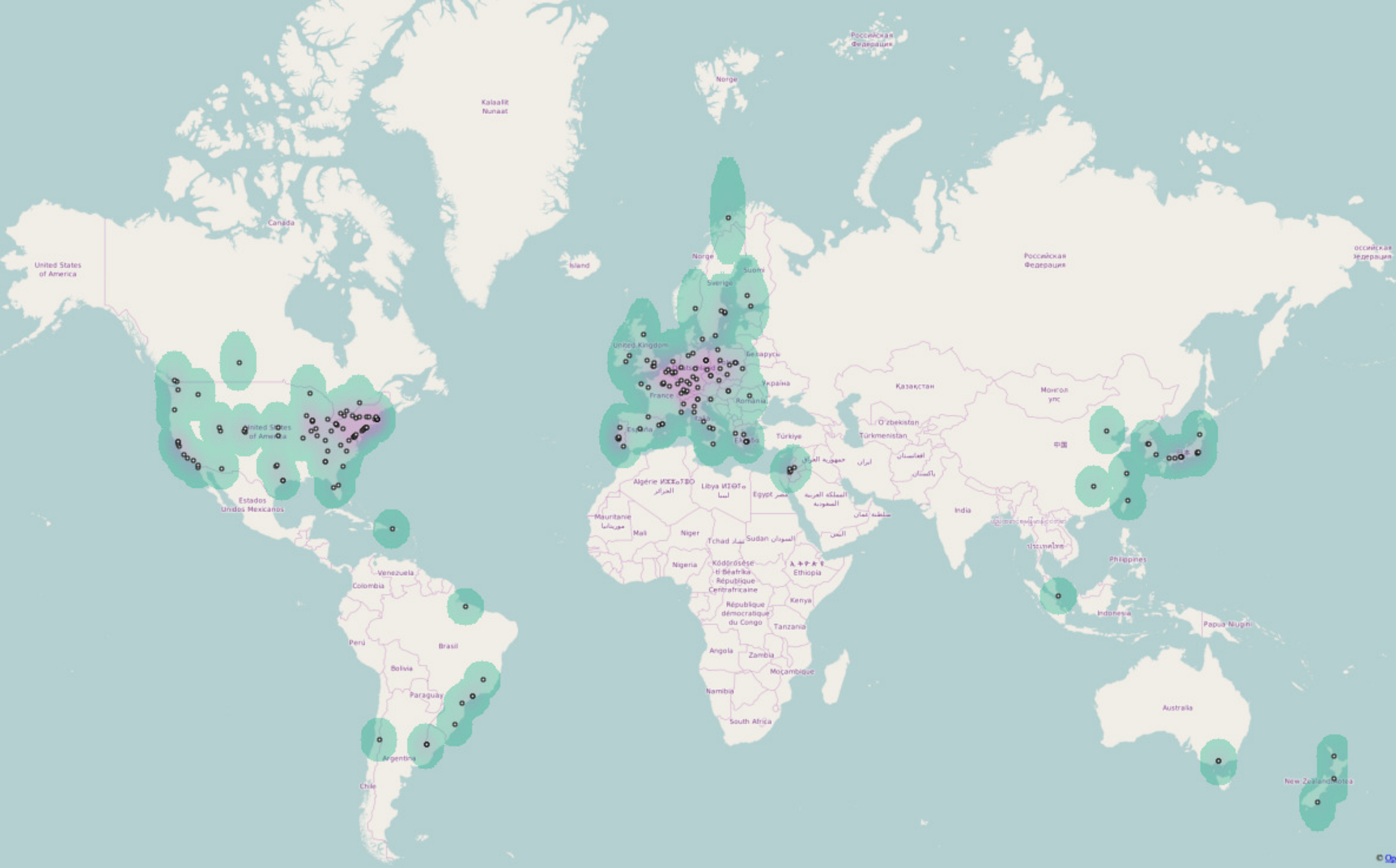}
\label{fig:world-targets} } 
\caption{Position and density of landmarks and targets.} 
\label{fig:worldmap} 
\end{figure*}

When remotely triggered, smartphones acquire their position using the GPS unit
(as they have to operate as landmarks). Then they collect a number of RTTs
towards a target (or a set of targets) as specified by the central server.
Probes are based on UDP since raw sockets, needed for generating ICMP messages,
are not available on Android devices without superuser privilege
level~\cite{6260474}.  The position of landmarks at the time of measurements is
shown in Figure~\ref{fig:world-landmarks}.

Approximately 400 localizations have been carried out using the nodes of
the PlanetLab network as targets. The real position of PlanetLab nodes is known,
this allowed us to compute the error between the estimated position and the
real one\footnote{A preliminary check has been carried out to remove
possible targets with clearly inaccurate position information.  In particular,
we eliminated from the list both those hosts were position was reported with
insufficient precision and those with invalid coordinates (some inconsistencies
have been pointed out in \cite{7296342}).}.  The placement of the target hosts
is shown in Figure~\ref{fig:world-targets}.

Note that both landmarks and targets are not uniformly distributed over the
globe.  In particular they are more densely deployed in North America and
Europe, whereas the other continents are not so well covered.  This reflects
the distribution of both PlanetLab nodes and Portolan participants. 

Measurements collected by smartphones are transmitted to a central server
where they are stored. Analysis has been carried out off-line to ensure
experimentation of different strategies and repeatability of results.

\begin{figure}[t] 
\centering
\includegraphics[width=1.0\columnwidth]{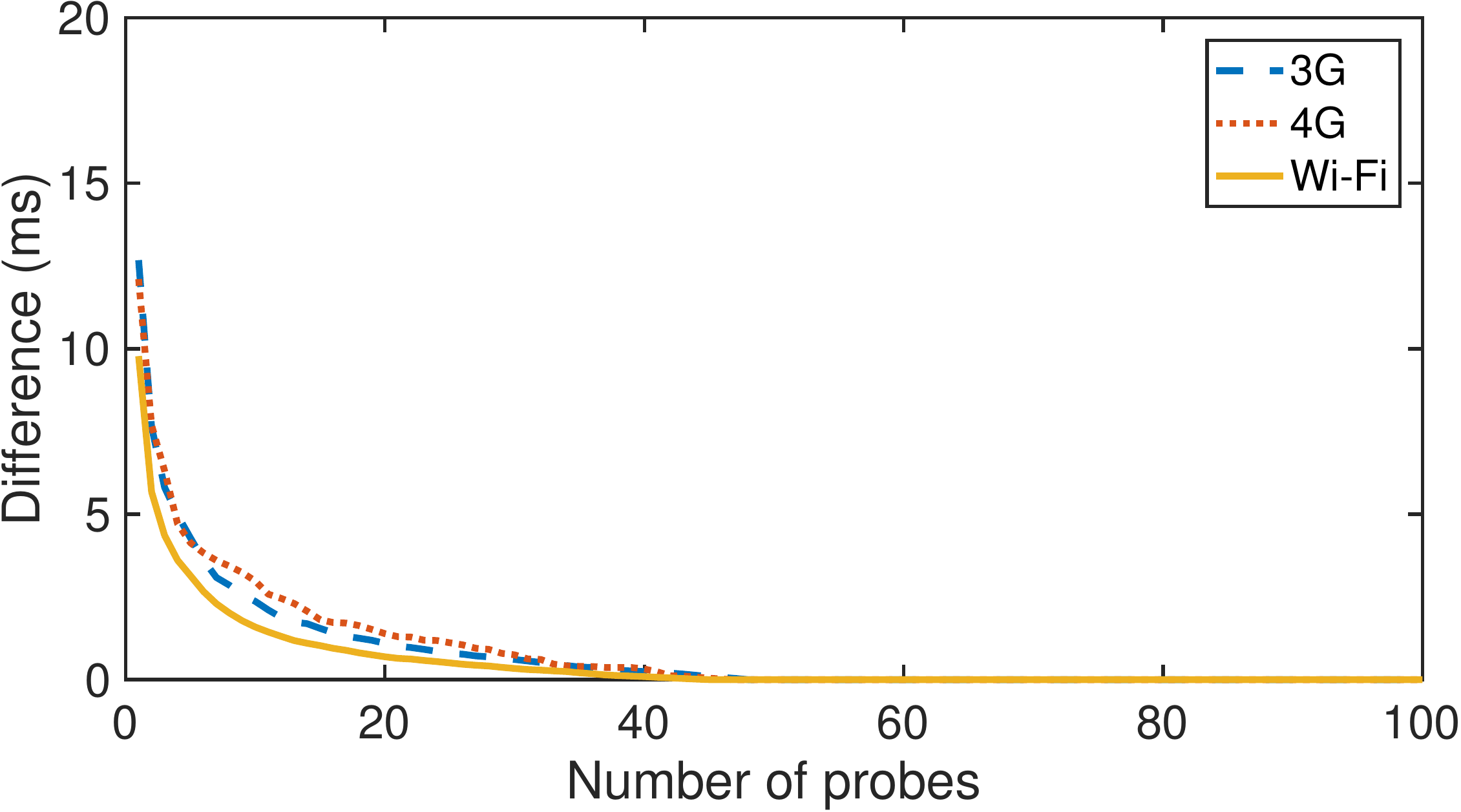} 
\caption{Stability of the minimum delay value when varying the number of probes.} 
\label{fig:convergence} 
\end{figure} 

To define the size of bursts -- the number of probes $K$ to be sent towards a
target -- we performed a preliminary analysis of delay variability.  First, we
collected a set of delay measurements using a number of bursts where each burst
was 100 probes long. For each burst we found the minimum delay when using all
the probes and when using the first $J$ probes (with $J \le 100$). We then
computed the difference between these minimum values, as an indicator of the
stability of the minimum delay when using bursts of different size. More
formally, given the $i$-th burst $\mathcal{M}_i = \{m_{i, 1}, m_{i, 2}, ...,
m_{i, 100}\}$, we first computed $\mathcal{M}_i^J = \{m_{i, 1}, m_{i, 2}, ...,
m_{i, J}\}$  for all $1 \le J \le 100$.  Then, we calculated $\mathcal{V}_i =
\{v_{i, 1}, v_{i, 2}, ..., v_{i, 100}\}$ where $v_{i, j} = min(\mathcal{M}_i^J)
- min(\mathcal{M}_i)$.  Figure~\ref{fig:convergence} shows the median
  $\mathcal{V}_i$ value derived from collected data (approximately 5800
bursts).  Results show that Wi-Fi connections are slightly more stable, from
this point of view, than 3G and 4G connections.  As expected, using bursts
composed by a large number of probes improves the stability of the minimum
observed delay. However, at the same time, the larger the burst the higher the
impact on the users' terminals (both in terms of generated traffic and energy
consumption). We set $K=50$ as a tradeoff between accuracy of measurements and
use of participants' resources.  In fact, during the data collection campaigns,
we noted an increase in the number of users who left the platform.  It is worth
to highlight that, as already mentioned, participation to the experiments was
not incentivized and it was based only on the will of users. Obviously, in case
of more favorable scenarios (e.g. in case of stricter control on terminals) the
use of longer bursts would lead to possible improvements in terms of
localization accuracy.

Smartphones are, by nature, mobile devices. Nevertheless, in the considered
scenario, they have to operate as landmarks, thus we introduced some mechanisms
aimed at limiting possible problems due to high mobility rates. The app
running on smartphones collects the position of the terminal both at the
beginning of a burst of probes and at the end.  In case of significant changes
(more than 5 km) the set of measurements is discarded.  Beside position, the
app also checks that the type of connection remains the same throughout the
transmission of the burst.  It is worth to note that a change in the type of
connection (e.g.  from cellular to Wi-Fi) can be a source of large
inconsistencies, probably bigger than the ones introduced by mobility. In fact,
a change in the type of connection may involve a completely different path
towards the target (for example because access takes place via a different
autonomous system). 

A smartphone can use both the GPS unit and network positioning to determine its
own location. Network positioning uses the visible cellular towers in the radio
range to locate the device, and it is characterized by an accuracy level worse
than GPS. We believe that the accuracy provided by network positioning is still
compatible with IP geolocation purposes.  However, the dataset used in our
analysis includes only measurements where the position of smartphones was 
computed using the GPS.  This was done to reduce the effects of unknown
factors on the proposed IP geolocation procedure.

\subsection{Difficulties due to crowdsourcing and wireless access}

The measurement collection phase presented some additional challenges, with
respect to previous studies, due to crowdsourcing and wireless access. 

\begin{figure}[t] 
\centering
\includegraphics[width=0.9\columnwidth]{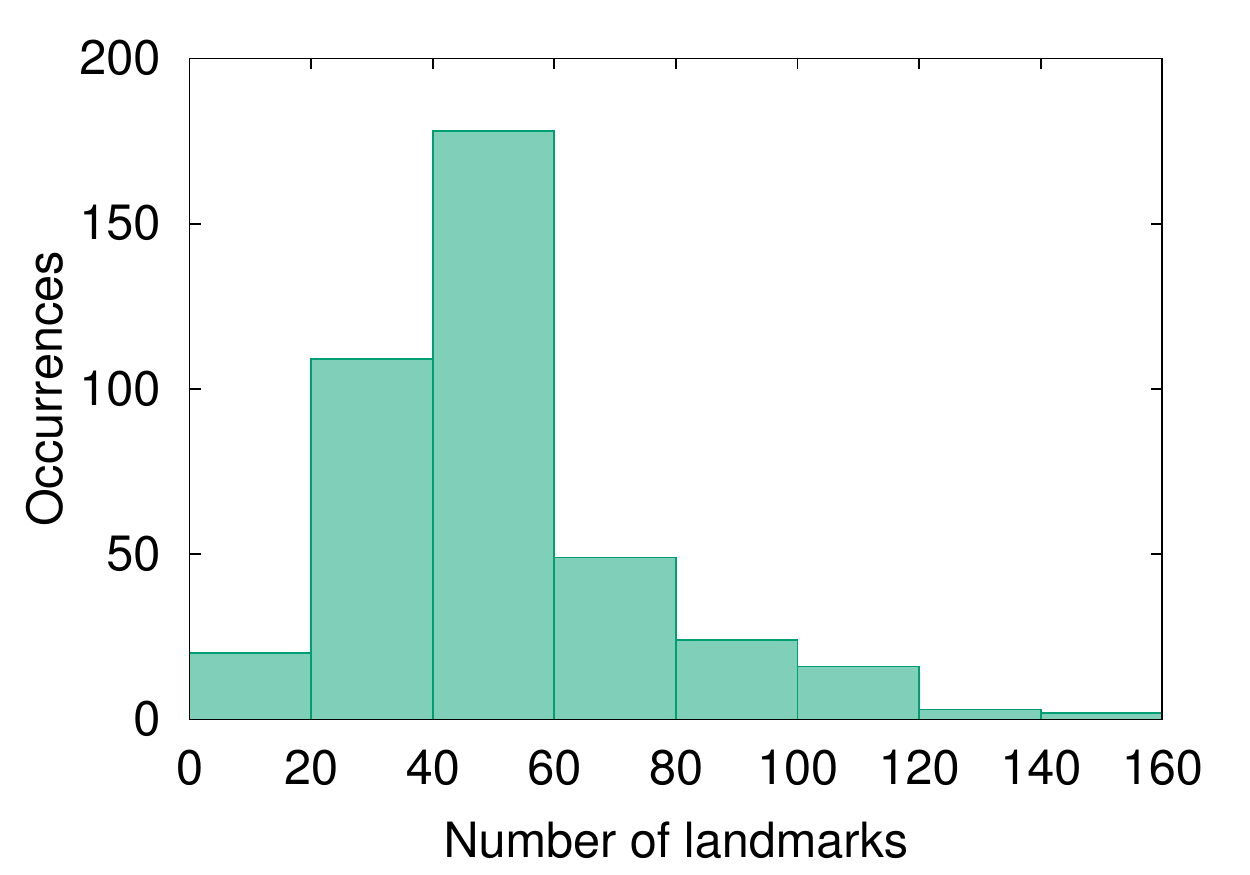} 
\caption{Number of landmarks involved in localizing a given target (average: 51.38, median: 48).} 
\label{fig:variability} 
\end{figure} 

A consequence of using a crowdsourcing-based approach is the high variability
in the number of smartphones enrolled in measurements. The number of devices
participating in localizing a single target varied from few units to $\sim150$.
Figure~\ref{fig:variability} shows this variability in more detail.  The reason
is due to the lack of control on devices, which are voluntarily contributed by
people participating in the Portolan platform. The user base is intrinsically
dynamical, as new users join the system while others terminate their
participation. Moreover, even if the number of enrolled users is relatively
constant in a given period, the availability of devices may still be subject to
rapid changes. For instance they may be turned off by their owners, run out of
battery, or enter a not covered area.

Operating conditions are also influenced by the wireless access of smartphones,
which makes communication more disconnection- and error-prone than when using
wired devices. Thus, some RTT measurements failed because of lost probes or
replies; similar problems occurred because of device mobility, as the device
may enter an area that is not covered by cellular networks or Wi-Fi access
points.  These difficulties arise not only when collecting RTT measurements,
but also when the central server communicates with enrolled devices for
coordinating their activities. In particular, the command used to trigger
measurements may be delayed because of a temporary disconnection of landmarks.
Because of problems occurred at runtime, the number of RTTs actually collected
by smartphones towards a single target was, on average, less than 50 (the chosen value of $K$). Problems
at runtime included disconnections, packet losses, filtering, and energy
exhaustion. In the end, the average number of samples collected by each
smartphone towards a given target was about 30. 

We roughly compared the variability of measurements collected using smartphones
with wireless access with respect to measurements collected using common PCs in
a wired-only scenario. In particular, we analyzed approximately 800 thousand
bursts collected using Portolan, and a similar number of bursts collected by
the PingER project~\cite{841837}. PingER is a measurement infrastructure aimed
at studying the end-to-end delay on the Internet. PingER comprises
approximately 39 probing machines and 430 probed hosts. In PingER, each
probing machine periodically sends a sequence of probes towards a set of
targets. Probes are based on ICMP. Collected RTTs are stored in a database and
are available to other researchers through a Web interface. PingER uses bursts
composed of ten probes. To ensure a fair comparison, we limited the analysis to
the first ten elements of each burst also for the measurements collected using
Portolan. First, for each burst, we selected the minimum delay value. Then we
computed the difference between the ten observed delays and the minimum. In
this way we obtained ten ``deviations'' from the minimum for each burst.
Finally we computed the mean value of deviations.  The dataset collected using
PingER has a mean deviation value of about 5 ms, whereas the one collected
using Portolan has a value of about 45 ms. It is worth to point out that the
difference between the two values originates from a number of factors including
not only the wireless/wired access and computational power, but also the
degree of heterogeneity, and the adoption of considerably different operating
systems and software layers.
The reader is forwarded to~\cite{7218402} for further details about 
variability of network delays in Android smartphones.

\section{Delay-distance model}
\label{lab:calibration}

\begin{figure}[t] 
\centering
\includegraphics[width=0.99\columnwidth]{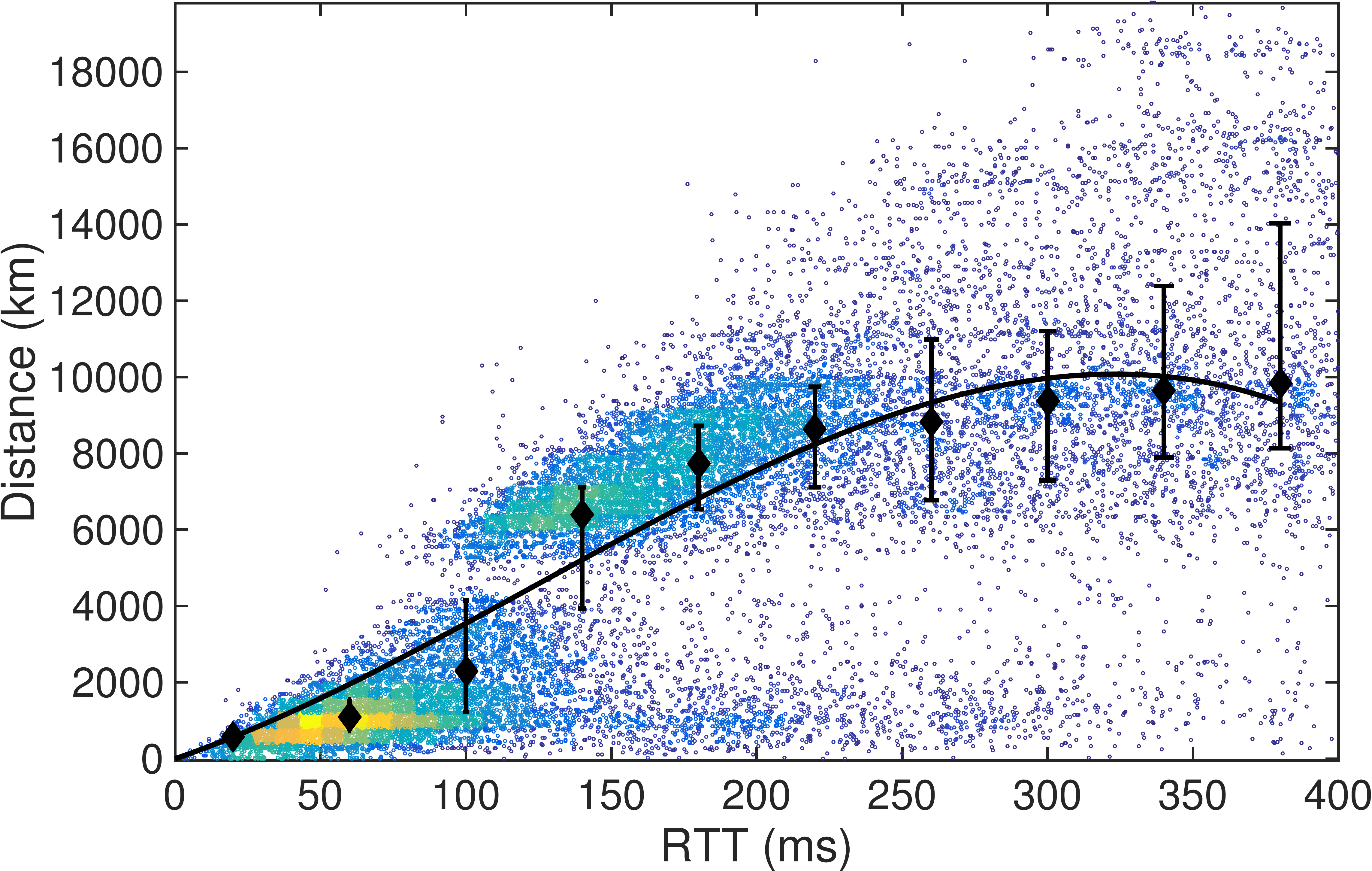}
\caption{Scatter plot of minimum RTT against great circle distance between landmark and target.}
\label{fig:scatterplot-all} 
\end{figure}

The end-to-end delay between two Internet hosts can be decomposed into these
terms: $i)$ transmission delay, i.e. the time needed to emit all the bits of the
packet; $ii)$ processing delay, i.e. the time required to process the packet;
$iii)$ queuing delay, i.e. the amount of time spent by a packet in queues before
being served; $iv)$ propagation delay, i.e. the time needed for signals to travel
from the transmitter to the receiver. These delays are introduced at every
link/router along the path. Some of these components are deterministic
(transmission and propagation delay), whereas others are stochastic (queuing
and processing delay). The only component that is related to geographical
distance is the propagation delay.  Since perturbations caused by queuing and
processing are always additive, their impact can be limited by measuring the
RTT a number of times and selecting the minimum value.
Circuitous routing is another source of inaccuracy, as the physical path
traversed by packets can be largely deviant with respect to the shortest
distance measured along the surface of the earth (the so called {\em great
circle distance}).

Using smartphones as landmarks also brings additional difficulties, as
previously explained. In particular, measurements are less reliable and
characterized by increased jitter with respect to when using more traditional
platforms. 

To determine the function $f()$ that represents the delay-distance model (used
in Equation~\ref{eq:radius}), we collected $\sim20750$ measurements where the real
distance between source and destination hosts is known (we used PlanetLab nodes
as targets, whereas the position of landmarks has been acquired using GPS).
The scatter plot of real distances against the $\hat m$ measured by landmarks is
shown in Figure~\ref{fig:scatterplot-all}. 

Data has been divided in bins. Median and standard deviation for each bin
are also shown in Figure~\ref{fig:scatterplot-all}. We performed a 
regression to find the polynomial model of the delay-distance relationship.  Such
relationship is, in theory, linear if we consider that end-to-end delay is the
sum of the four components mentioned above (the only component related with
distance is the propagation delay, which linearly increases with distance).
However, as the delay gets larger, other factors have an influence on the
relationship, that thus becomes better represented by a polynomial function
($f()$ is depicted by the curve in Figure~\ref{fig:scatterplot-all}).
For instance, very large round trip times may be produced by non-empty
queues found in routers along the path (as the distance increases, the number
of traversed routers increases as well). The use of a polynomial function for
converting delays into distances has been suggested also
in other works (e.g. \cite{laki11:spotter,Dong201285}). 

\begin{figure}[t] 
\centering 
\subfloat[Same continent]{
\includegraphics[width=0.99\columnwidth]{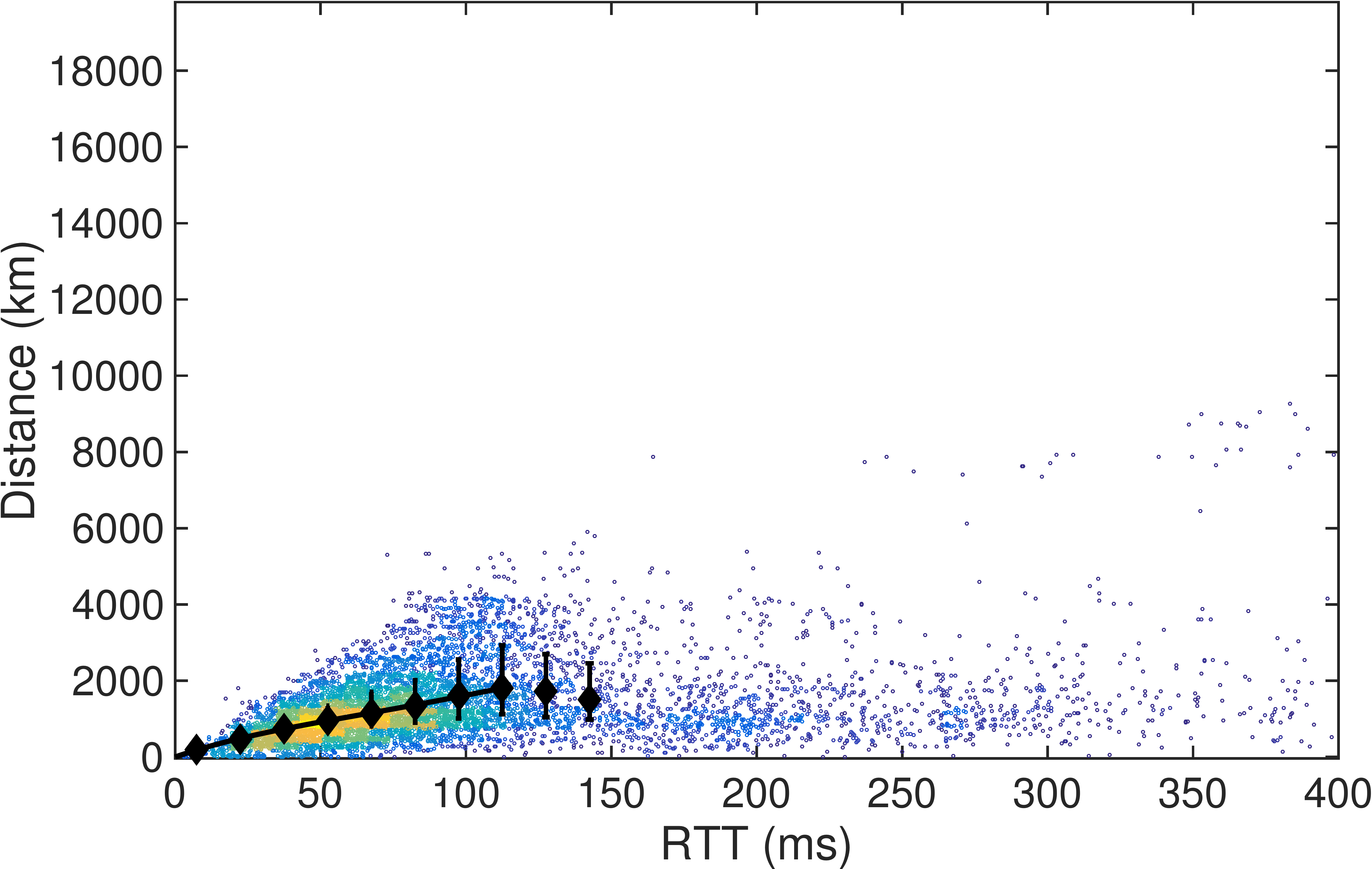}
\label{fig:same-continent} } 
\hspace{1cm}
\subfloat[Different continents]{
\includegraphics[width=0.99\columnwidth]{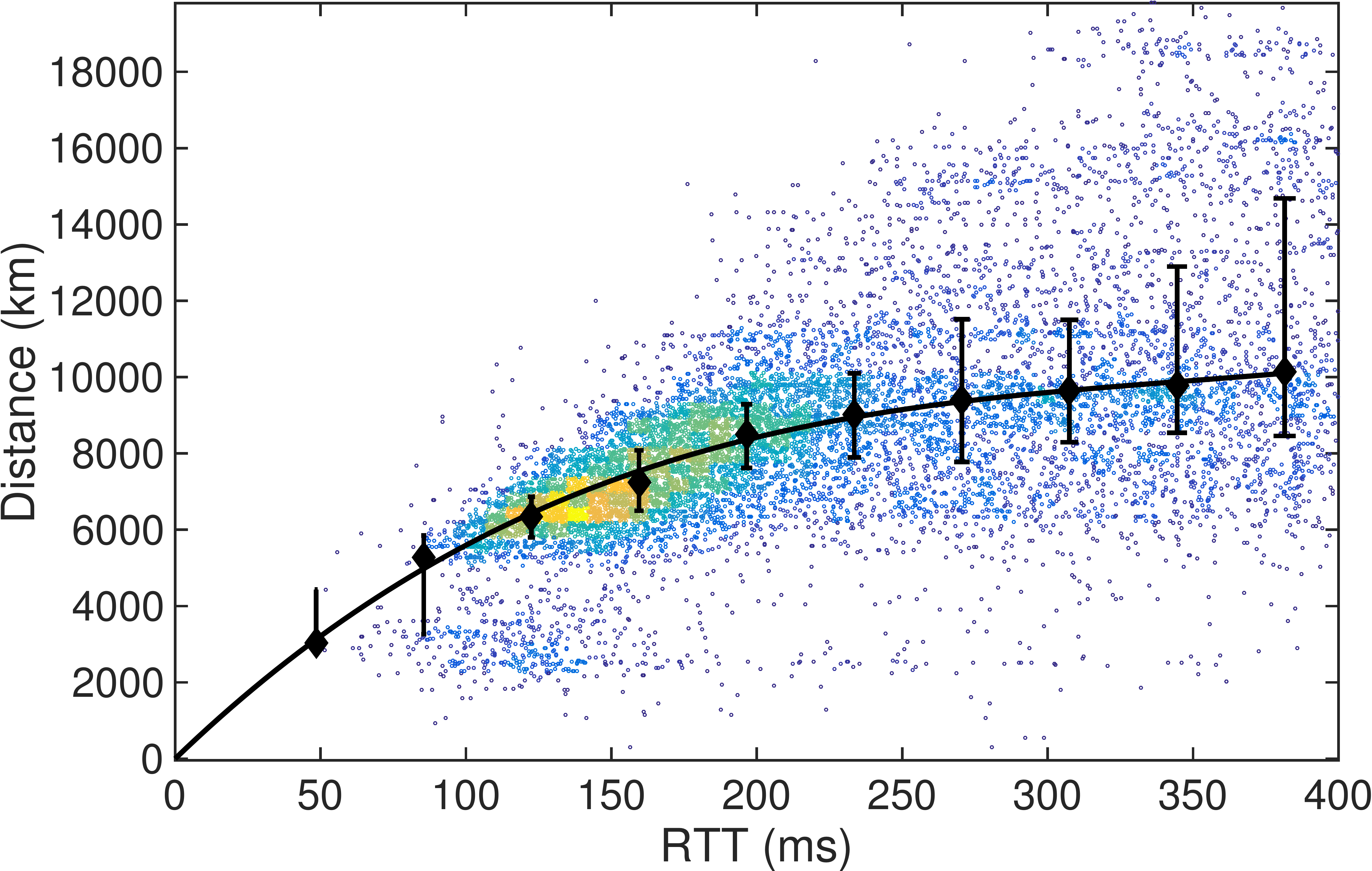}
\label{fig:different-continent} } 
\caption{Measurements with target and landmark located in the same continent or in different continents.} 
\label{fig:same-different} 
\end{figure}

By analyzing Figure~\ref{fig:scatterplot-all}, some considerations can be made.
First, the standard deviation of distances for a given delay value increases
with the delay itself. This means the further away from the target a landmark
is, the less reliable the measurement is.  Second, two major agglomerations of
points are clearly visible.  Inspection of data suggested that the
presence/absence of long-haul links connecting different continents is
significant from this point of view.  Figure~\ref{fig:same-continent} includes
only measurements where both target and landmark belong to the same continent.
Figure~\ref{fig:different-continent}, conversely, includes only measurements
where target and landmark belong to different continents. 

The clear separation between the two sets of measurements suggests the use of
two different functions for converting delays into distances.  Let us call
$f_{s}()$ and $f_{d}()$ the polynomial functions obtained through regression of
measurements where target and destination are within the same continent or in
different continents respectively. The two functions are represented by the 
curves in Figures~\ref{fig:same-continent} and~\ref{fig:different-continent}.
The use of two different functions should be able to provide better estimations
with respect to the use of a single global model.  This solution is 
evaluated in Section~\ref{sec:results}.  Note that the use of two different
conversion functions requires additional knowledge: the system has to know if
source and destination are in the same continent or in different ones, as it
has to use either $f_s()$ or $f_d()$. Such information can be inferred from the
autonomous systems the two hosts belong to (with some exceptions, e.g. when an
autonomous system is spread over different continents).  In any case, the
localization procedure becomes more complicated and requires additional
information (a mapping from autonomous systems to continents).

Several previous works relied on landmark-specific calibration, i.e. a
different delay-distance model is used at every landmark
(\cite{Gueye:2006:CGI:1217687.1217693, Wong:2007:OCF:1973430.1973453}). This
approach is supposed to increase localization accuracy, as it is able to capture
the topological peculiarities of every landmark. Other authors, on the
contrary, supported the use of a unique model for all landmarks
(e.g~\cite{laki11:spotter}).  As previously stated, in our scenario characterized by
the use of smartphones, the adoption of landmark-specific calibration does not
makes much sense. 

\begin{figure}[h!] 
\centering 
\subfloat[3G]{
\includegraphics[width=0.99\columnwidth]{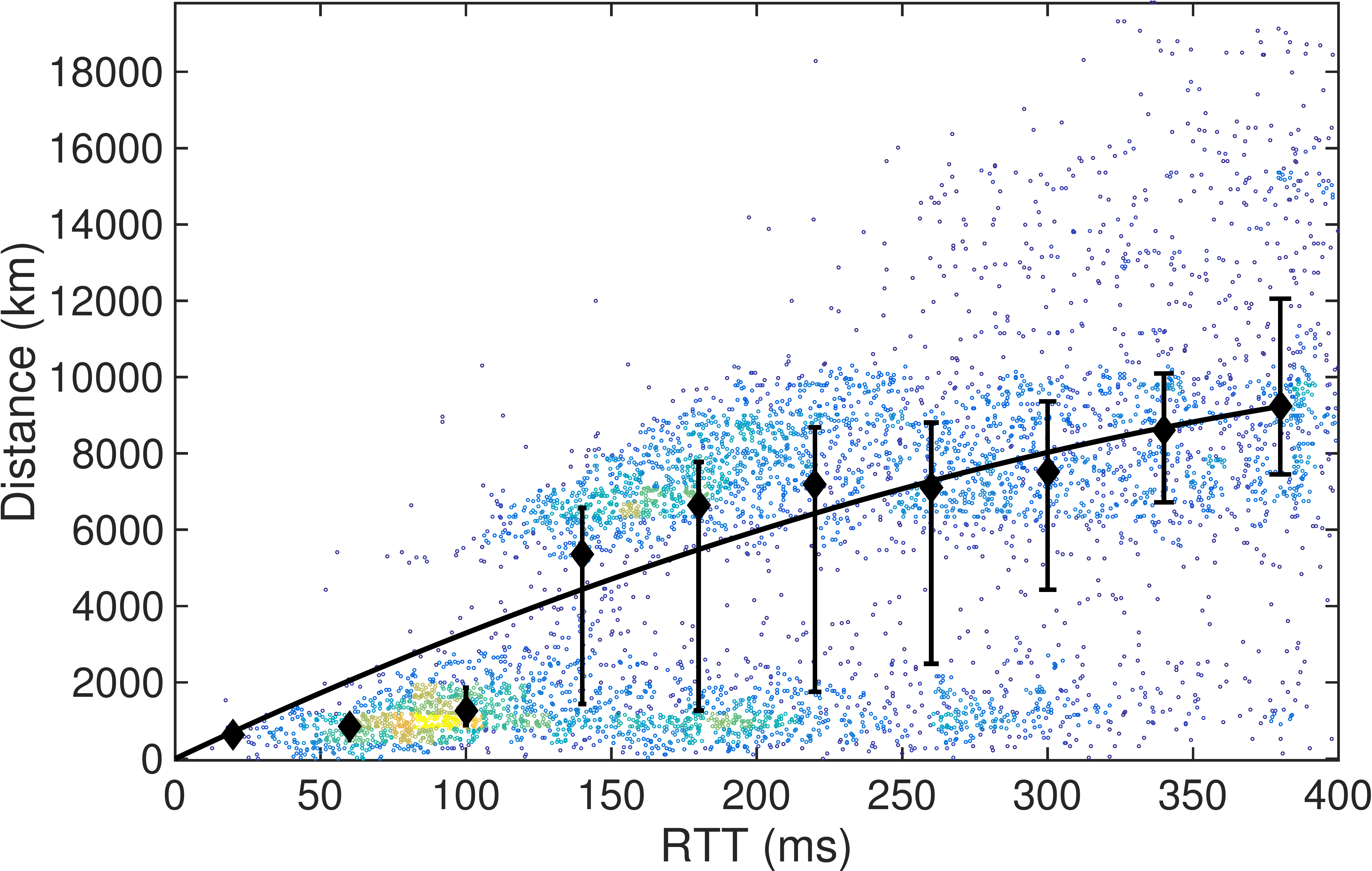}
\label{fig:distance-rtt-3g} } 

\subfloat[4G]{
\includegraphics[width=0.99\columnwidth]{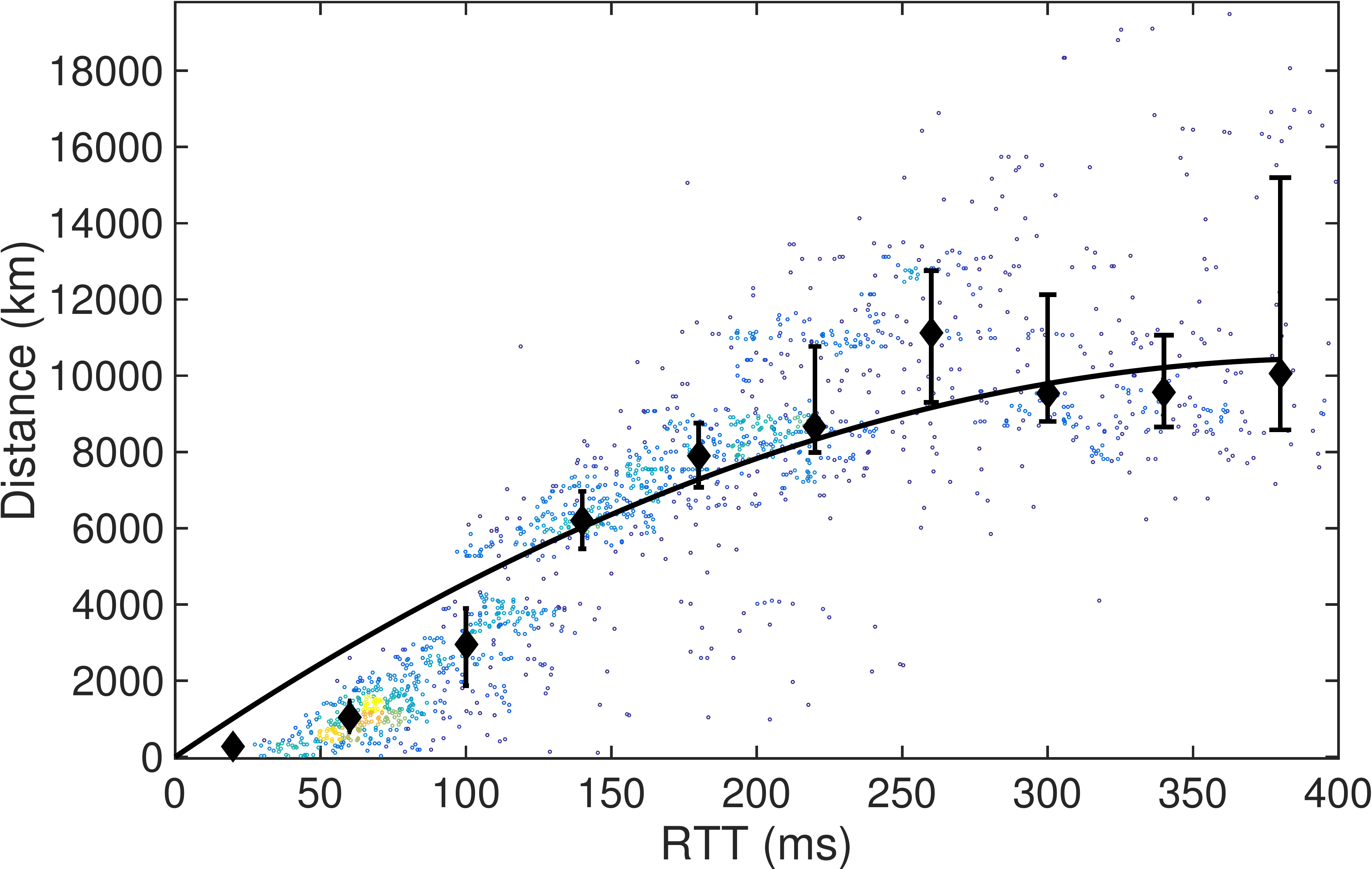}
\label{fig:distance-rtt-4g} }

\subfloat[Wi-Fi]{
\includegraphics[width=0.99\columnwidth]{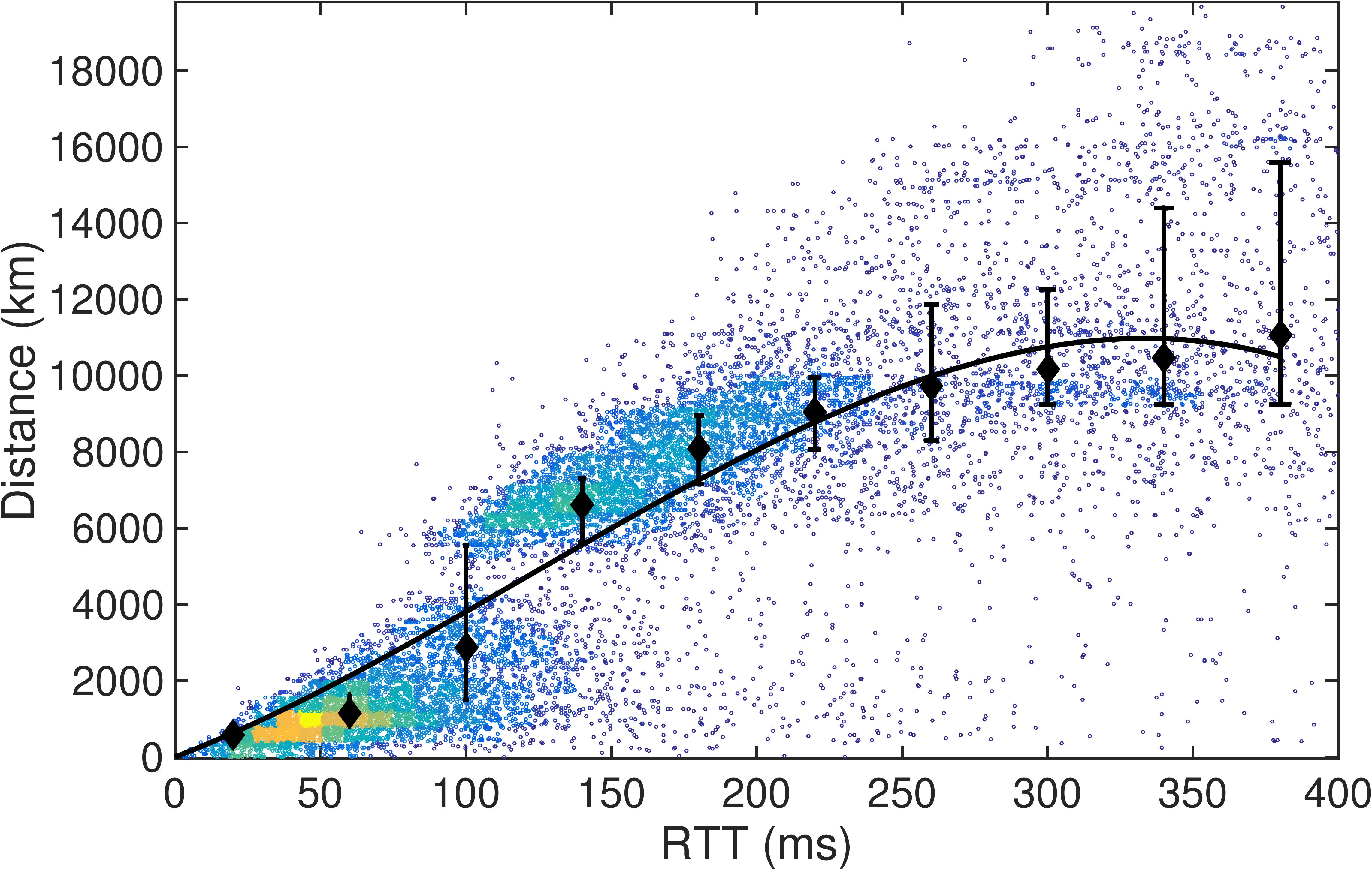}
\label{fig:distance-rtt-wifi} } 
\caption{Distance against observed minimum delay for the different access technologies.} 
\label{fig:compare-rtt-net-types} 
\end{figure}

\begin{table} 
\centering 
\caption{Access technologies and percentages of measurements.}
\label{tab:access_type}
\begin{tabular}{lr} 
\hline 
\textbf{Type} & \textbf{Measurements (\%)}\\ 
\hline 
3G &  $\sim28\%$ \\ 
4G & $\sim8\%$  \\ 
Wi-Fi & $\sim60\%$ \\ 
Other/Not available & $\sim4\%$ \\ 
\hline 
\end{tabular}
\end{table}

Even though per-landmark calibration is not feasible, we considered that
possible improvements may be achieved through the development of delay-distance
models that are technology-specific. In fact, every smartphone is able to
understand which is the type of access technology currently in use, and this
information can be reported to the central server together with delay
measurements.  Figure~\ref{fig:compare-rtt-net-types} shows a set of scatter
plots, one for each family of access technologies (3G, 4G, Wi-Fi).  For each
group of measurements a different $f()$ is obtained through regression,
similarly to the same/different continent criterion.  Let us call $f_{3G}()$,
$f_{4G}()$, and $f_{Wi-Fi}()$ the resulting functions. At runtime, the system
uses $f_{3G}$, $f_{4G}()$, or $f_{Wi-Fi}()$ depending on the technology
currently in use.  The percentages of samples for the different access
technologies, in our dataset, are reported in Table~\ref{tab:access_type}.  The
majority of measurements have been collected using Wi-Fi, followed by 3G, and
4G.  Few measurements have been collected using 2G and 2.5G cellular networks.
These measurements were not used because such communication technologies are
rapidly becoming obsolete. In some cases, smartphones have not been able to
report the exact type of access technology.  Also these measurements were not
used in our analysis.  The fraction of measurements collected using obsolete
cellular networks, or without indication of the technology used, is reported
as ``Other/Not Available'' in Table~\ref{tab:access_type}. 

In summary, we devised three different delay-distance models: $i)$ a model
obtained from the whole dataset (let us call this one the global model); $ii)$ a
model that takes into account whether the two hosts are in the same continent or
not (let us call this one the continent-based model); $iii)$ a model that
leverages information concerning the wireless access technology (let us call
this one the technology-based model). A fourth model is obtained by combining
the continent- and technology-based model, as they  exploit disjoint properties
(let us call this one the hybrid model).

Note that for large delay values the relationship between delay and distance
becomes practically meaningless. This is particularly evident in
Figure~\ref{fig:same-continent} where the two-rightmost median values do not
preserve the increasing relationship between delay and distance. Thus, we
decided to limit the use of $f()$ to the region where distance is positively
proportional to delay.

\section{Results}
\label{sec:results}

\begin{figure}[t] 
\centering
\includegraphics[width=0.99\columnwidth]{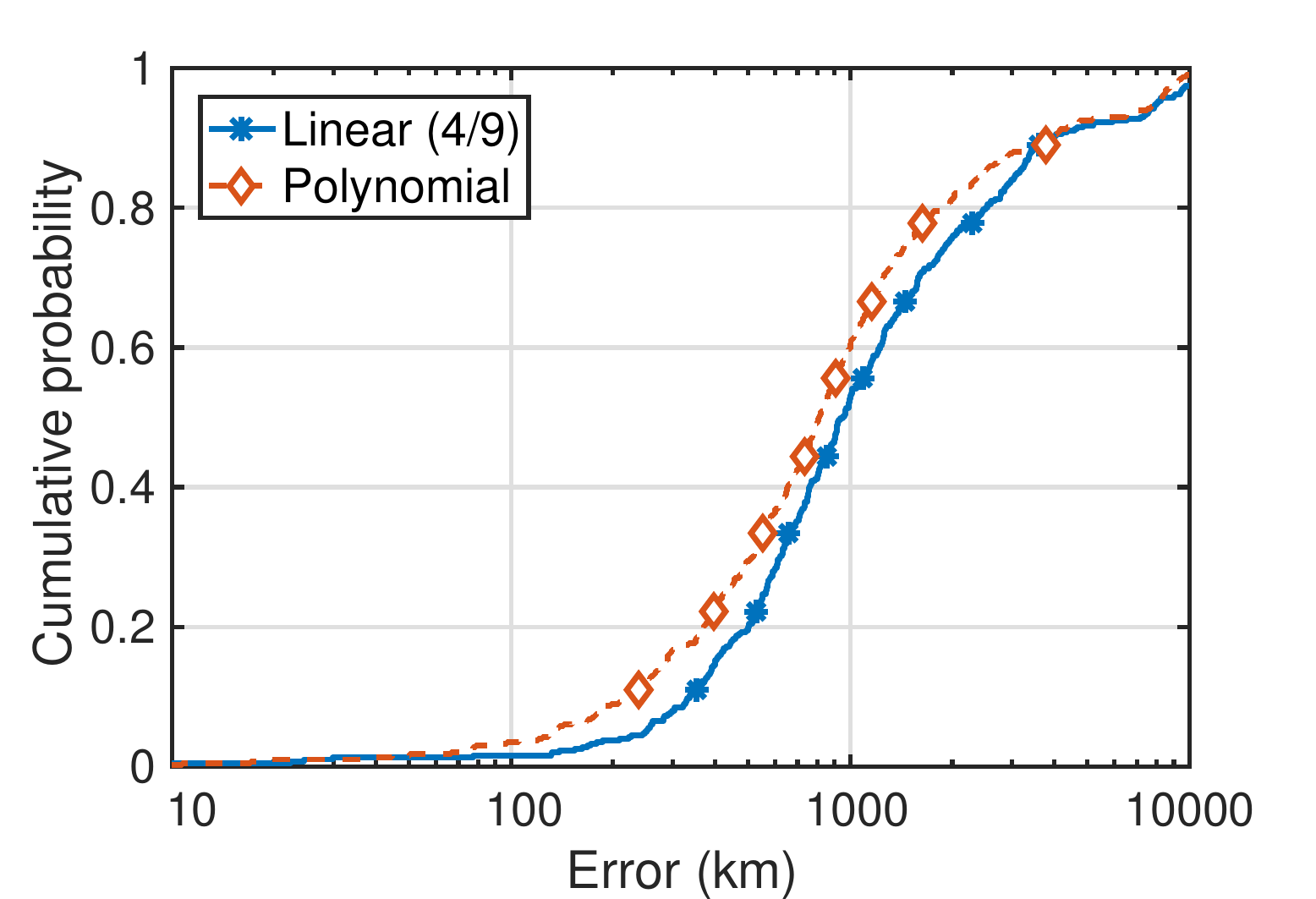}
\caption{CDF of localization error when using the global polynomial model 
and a linear one (the latter with coefficient 4/9, as suggested in~\cite{Katz-Bassett:2006:TIG:1177080.1177090}).}
\label{fig:cdf_compare} 
\end{figure}

This section reports the results obtained using the localization procedure in
combination with the different delay-distance models. To avoid using the same
data for both calibration and evaluation, we followed an approach based on
ten-fold cross-validation~\cite{Duda:2000:PC:954544}. In particular, we
separated the set of hosts used for deriving the delay-distance model from the
ones used for evaluation. With $k$-fold cross-validation data is partitioned
in $k$ disjoint subsets of equal size. Then $k - 1$ subsets are used to train the
system (in our case to perform model calibration) whereas the remaining subset
is used to evaluate the performance of the system (on previously unseen data).
The procedure is repeated $k$ times so that all subsets are left out during
training and used exactly once for evaluation. Results are finally aggregated
to obtain the overall performance of the proposed IP geolocation method.

Figure~\ref{fig:cdf_compare} shows the cumulative distribution function of
localization error when using the global delay-distance model (polynomial) and
a linear model with coefficient (4/9) derived from
literature~\cite{Katz-Bassett:2006:TIG:1177080.1177090}. Median localization
errors are respectively equal to $\sim 808$ km and $\sim 953$ km.  It is
evident that a linear model, calibrated for wired access networks, is less
adequate for a smartphone-based scenario. 

\begin{figure}[t] 
\centering
\includegraphics[width=0.99\columnwidth]{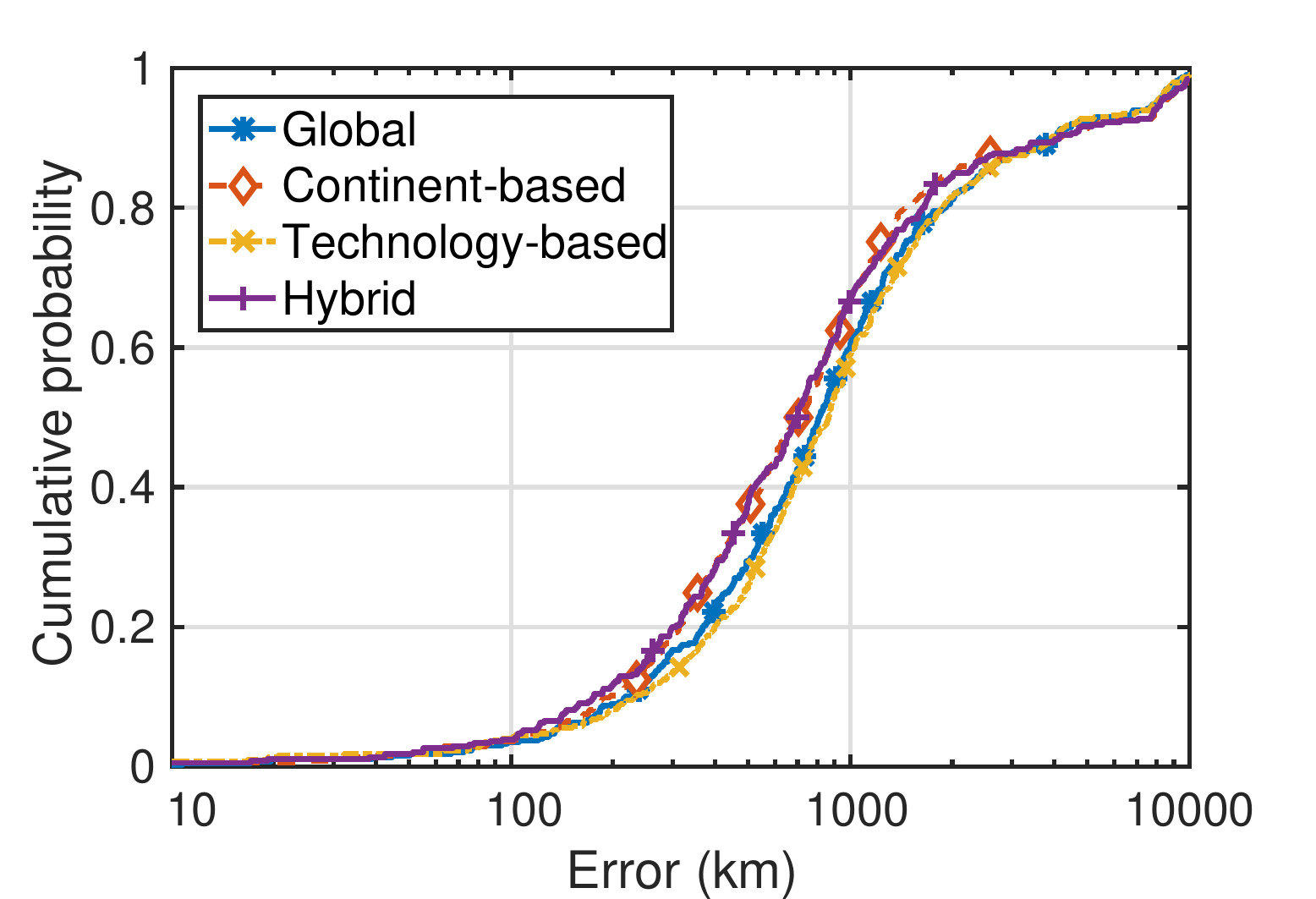}
\caption{CDF of localization error when using the different delay-distance models.}
\label{fig:cdf_different_models} 
\end{figure}

We then evaluated the performance of the continent-based model, the
technology-based model, and the hybrid model with respect to the global model.
Results are shown in Figure~\ref{fig:cdf_different_models}. The best results
are obtained when using the continental and the hybrid models. At this level,
the benefits introduced by using a model that depends on the access technology
are almost negligible.  

\begin{figure}[t] 
\centering
\includegraphics[width=0.99\columnwidth]{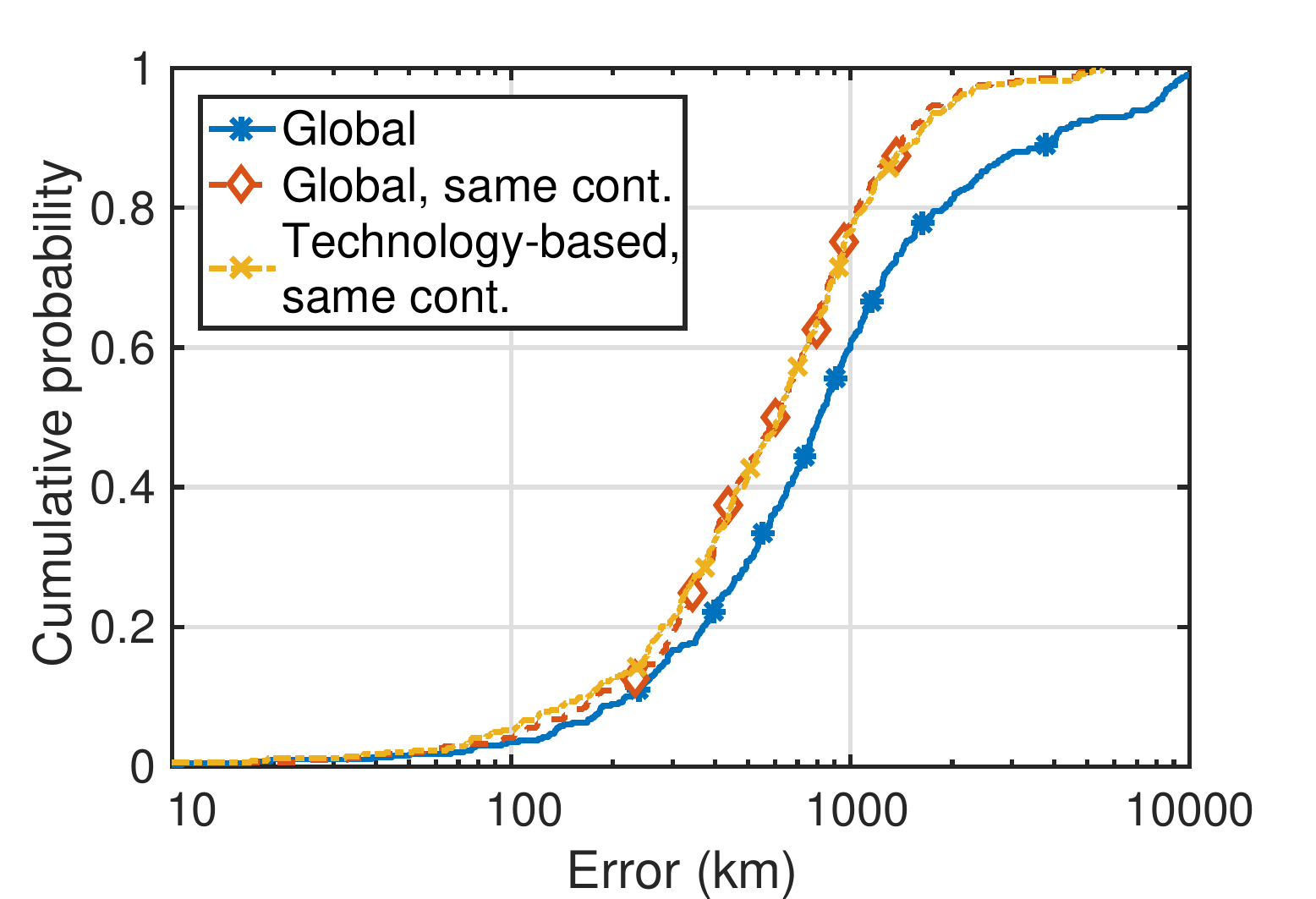}
\caption{CDF of localization error when limiting landmarks to the ones placed in the same continent of the target.}
\label{fig:cdf_errors_samecont} 
\end{figure}

Figure~\ref{fig:cdf_errors_samecont} shows the results achieved when limiting
the landmarks to the ones that are placed in the same continent of the target.
More precisely, when localizing a target a number of landmarks are located in
its same continent, whereas others are located in other continents.  We
discarded the measurements collected by the latter ones.  This produced an
improvement of localization accuracy. This behavior is somehow expected, as
distant landmarks provide less accurate measurements than close ones.  In
particular, in this scenario, the global model achieves a median localization
error equal to $\sim600$ km, while the technology-based model achieves a median
error equal to $\sim616$ km.

\begin{figure*}[t]
\centering
\subfloat[Global delay-distance model]{
  \includegraphics[width=0.45\textwidth]{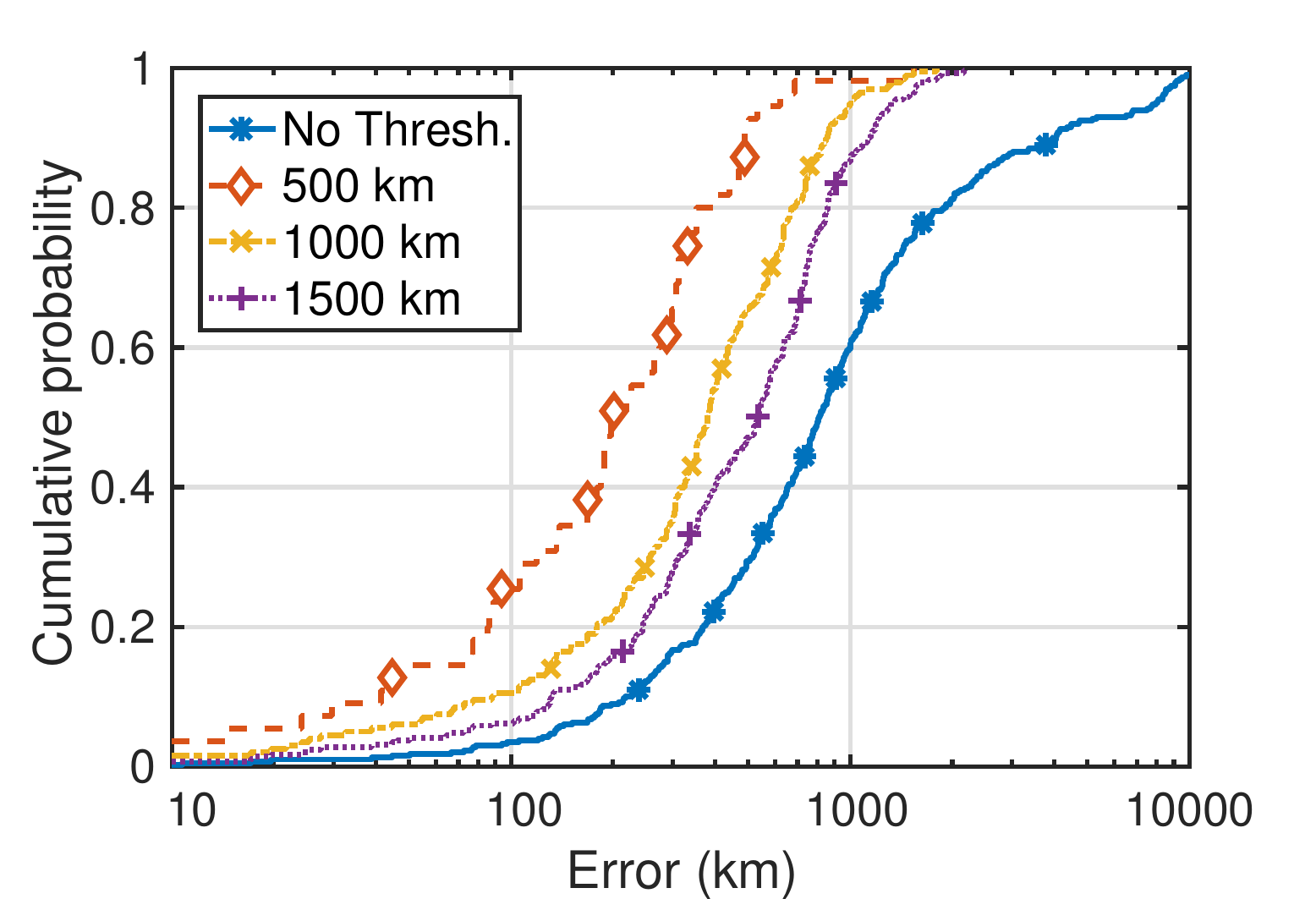}
  \label{fig:filtering_1} 
} 
\hfill 
\subfloat[Technology-based delay-distance model]{
  \includegraphics[width=0.45\textwidth]{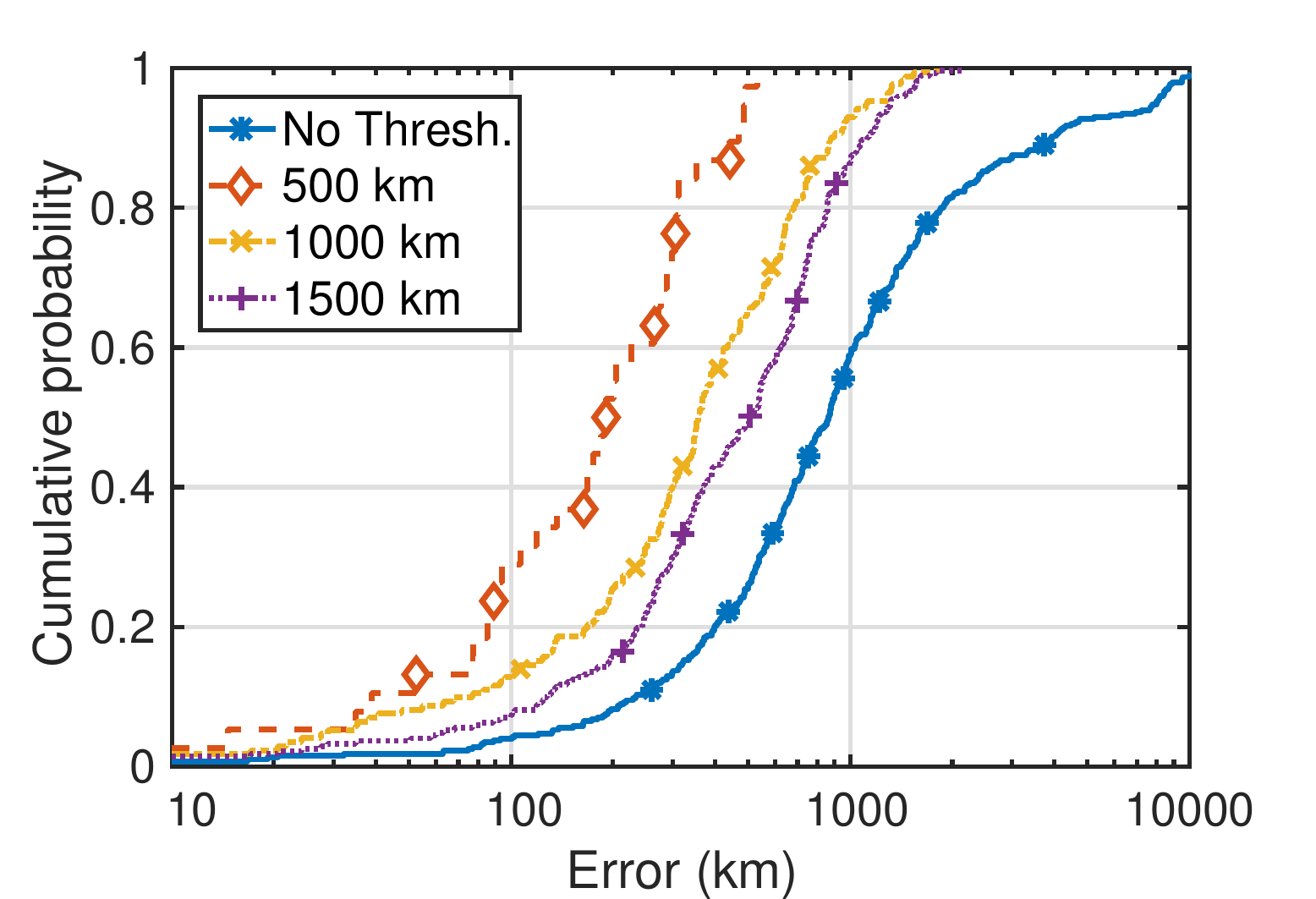}
  \label{fig:filtering_2} 
}
\caption{CDFs of localization error when filtering out distant landmarks (thresholds at 500, 1000, 1500 km).}
\label{fig:filtering}
\end{figure*}

Consequently, we studied the performance of the localization process when
filtering out distant landmarks. Figure~\ref{fig:filtering} shows the
cumulative distribution function of localization error when using only
landmarks within 500, 1000, and 1500 km from the target\footnote{A measurement
provided by a landmark is discarded if the RTT corresponds to an estimated
distance greater than the threshold.}. Results achieved without filtering are
also reported as a reference. It is evident that localization accuracy
increases when adopting more aggressive filtering.  In more detail,
Figure~\ref{fig:filtering_1} shows the results obtained when filtering is
applied to a localization procedure based on the global delay-distance model:
the median localization error is equal to 201, 379, 535 km when using the three
thresholds. Figure~\ref{fig:filtering_2} shows the results obtained when the
technology-based model is used. In this case the median errors are respectively
equal to 189, 358, and 507 km.

It is worthwhile to note that when filtering is applied, the technology-based
model is able to provide some benefits in terms of localization accuracy (about
6\%): when RTTs are small, the technology-based model is able to better capture
the relationship between delay and distance. 

\begin{figure}[t!] 
\centering
\includegraphics[width=0.99\columnwidth]{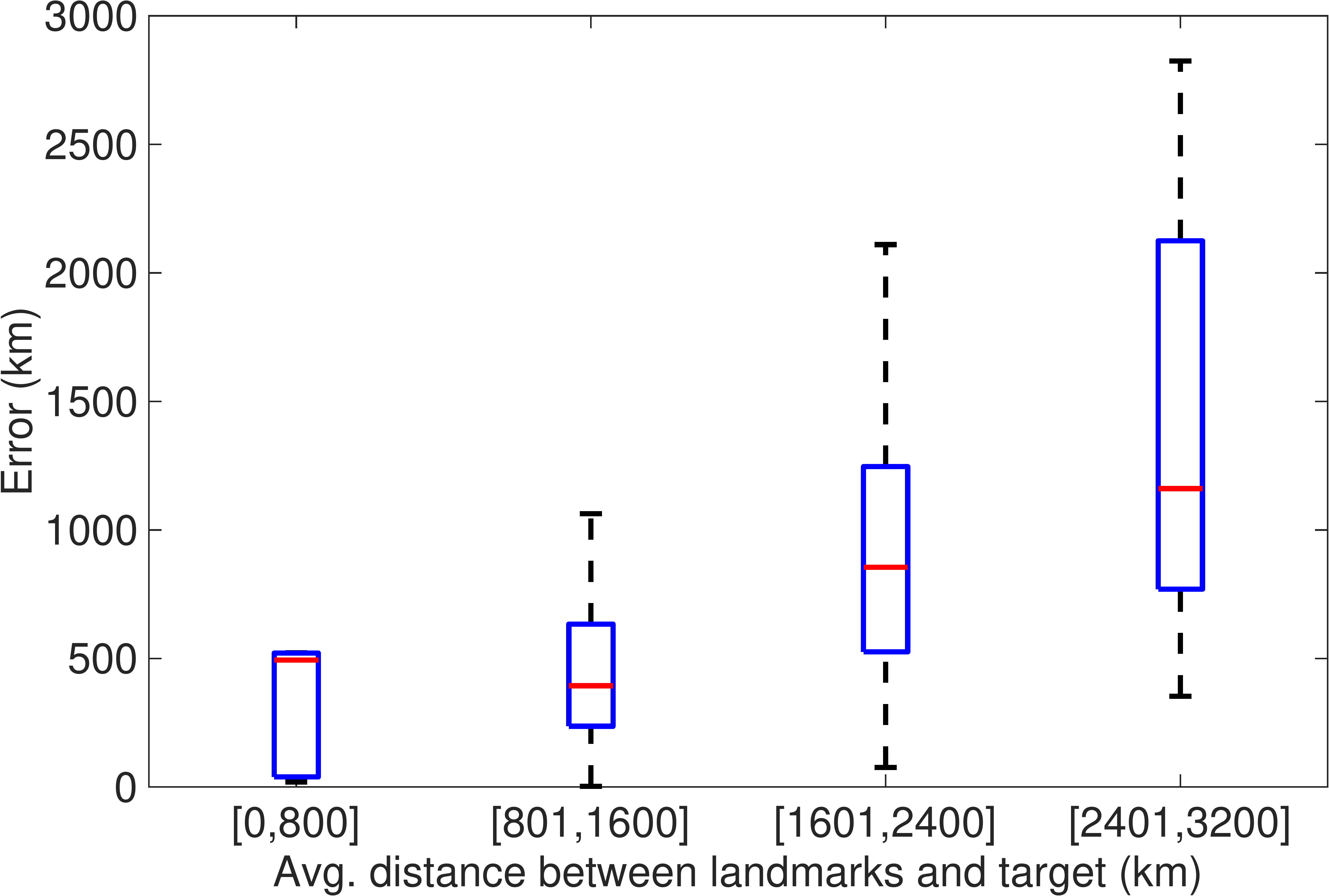}
\caption{ 
Localization error when varying the average target-landmark distance.}
\label{fig:error-landmarks-distance} 
\end{figure}

In any case, the larger the distance between target and landmarks the larger
the localization error (for instance because of queues or congestions).
Figure~\ref{fig:error-landmarks-distance} depicts this finding in terms of
localization error against average target-landmark distance.

\begin{figure}[t!] 
\centering
\includegraphics[width=0.99\columnwidth]{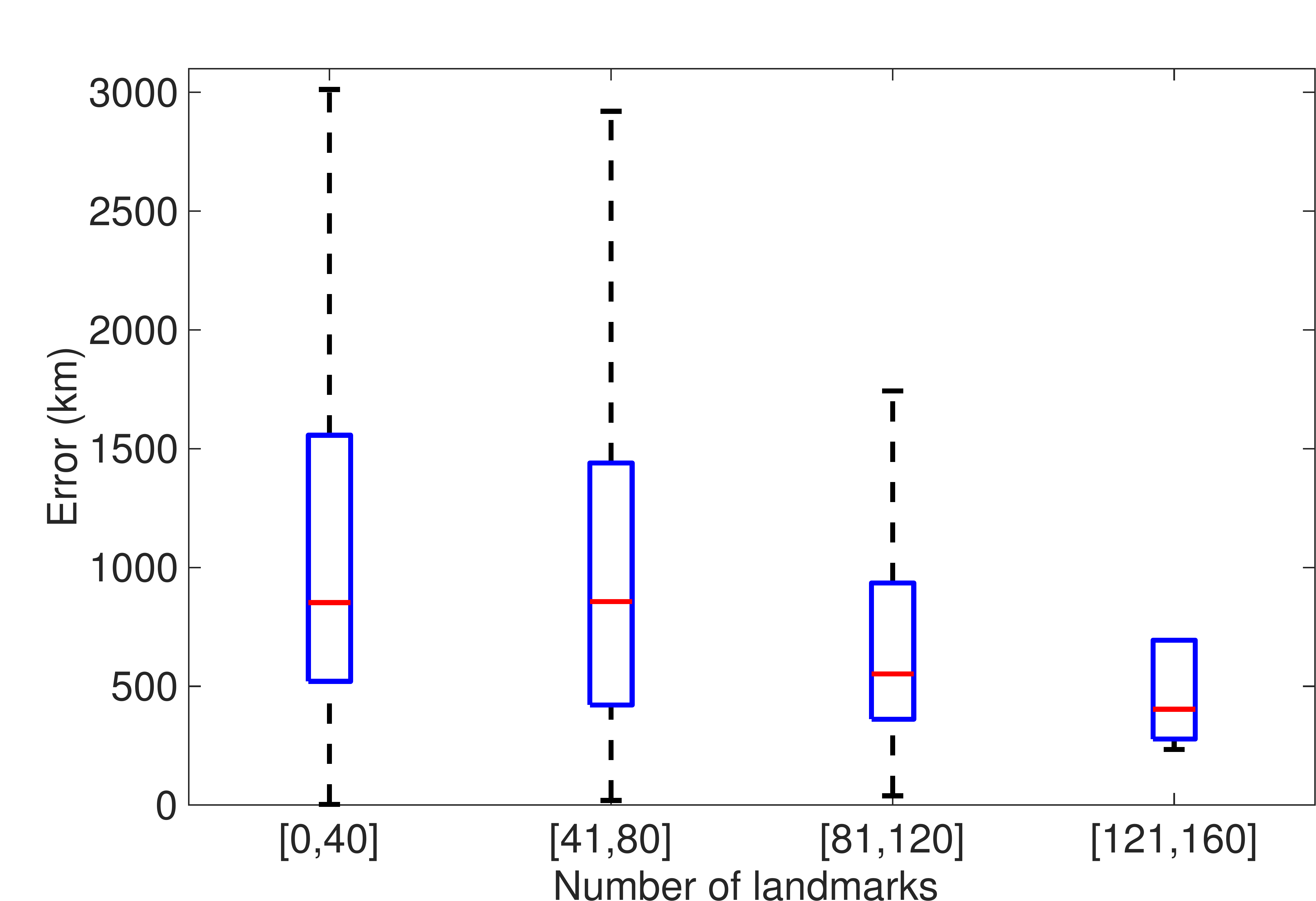}
\caption{ 
Localization error against the number of landmarks involved in localizing a target.
\label{fig:error-landmarks-ok} 
} 
\end{figure}

Another important factor affecting the performance of IP geolocation is the
number of landmarks that participate in locating a target.  Using a larger
number of landmarks generally provides better results, as shown in
Figure~\ref{fig:error-landmarks-ok}.  Thus, the performance of the system could
be improved simply by increasing the number of participants, as this would reduce
the average distance to the target and, at the same time, it would increase the
average number of landmarks involved in locating a target.  Since the proposed
method is based on crowdsourcing, increasing the number of devices is not
straightforward as it is not under direct control of
experimenters~\cite{Faggiani:2013:LLD:2536714.2536717}.  It is known that
participation to crowdsourcing systems can be stimulated through proper
incentives ~\cite{horton2010labor,Yang:2012:CSI:2348543.2348567,
mao_volunteering_2013}, but this possibility has not been put into practice in
current work and it is left for further studies.

\subsection{Comparison with closest landmark}

\begin{figure}[t!] 
\centering
\includegraphics[width=0.99\columnwidth]{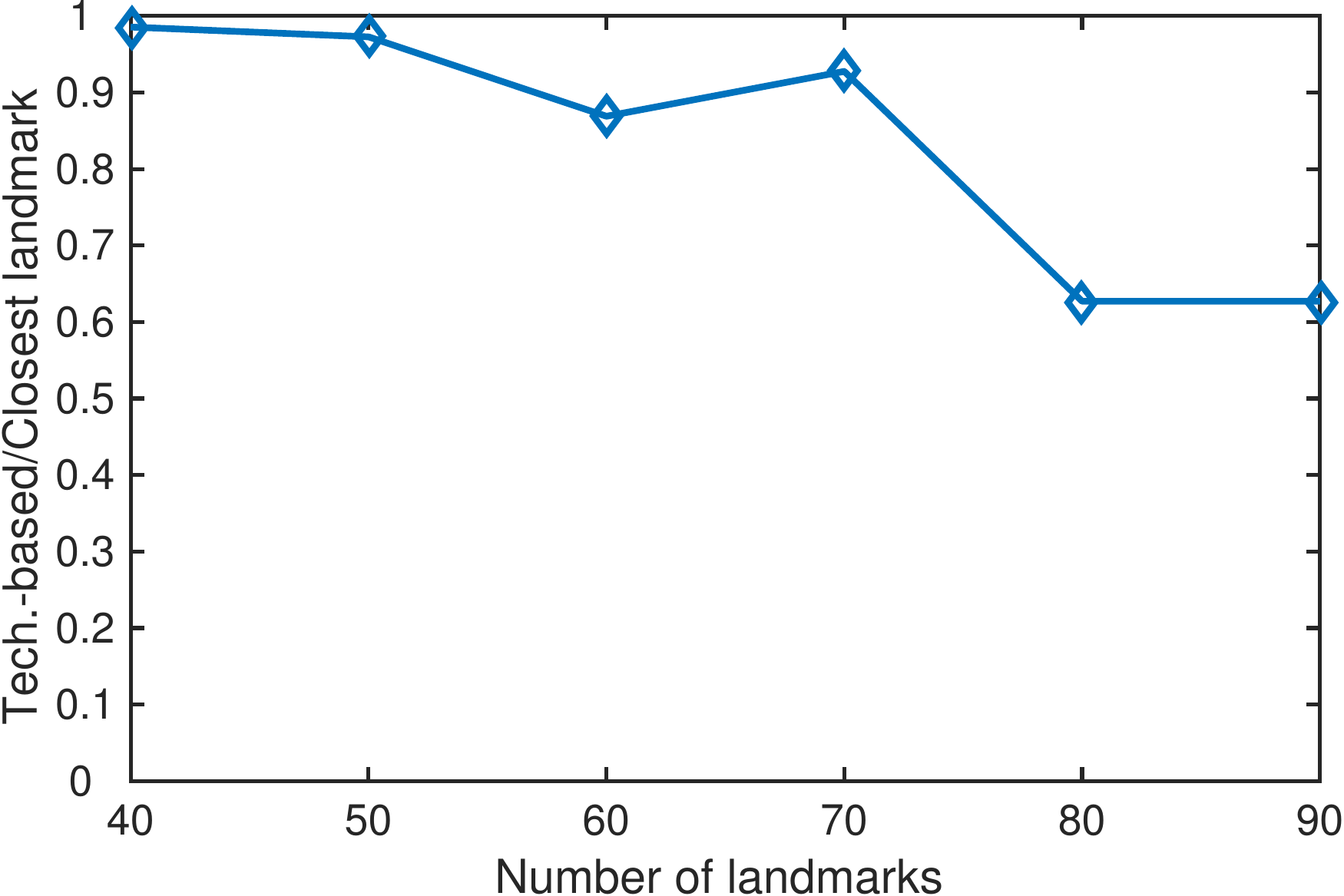}
\caption{ 
Ratio between the median error obtained by our technique and the median error obtained by closest landmark.
\label{fig:comparison-closest-minor} 
} 
\end{figure}

We compared the performance of the proposed method with respect to a simple
benchmark technique based on the closest landmark. In particular, such
reference method just returns, for each target, the coordinates of the closest
landmark in the delay space.  

Figure~\ref{fig:comparison-closest-minor} shows the ratio between the median
error obtained by the proposed method (technology-based) and the median error
obtained by closest landmark. The errors for the two methods have been computed
when varying the number of landmarks involved in localization. When the number
of landmarks is small, the performance of the two methods tend to be similar.
On the contrary, as the number of landmarks involved in the localization
process increases, the performance of the proposed method, with respect to
closest landmark, increases as well.  In fact, when the number of landmarks is
large the intersection of the regions defined by the different landmarks
becomes smaller, and this improves the localization process (this phenomenon is
also highlighted by Figure~\ref{fig:error-landmarks-ok}).

\section{Conclusion}
\label{lab:conclusions}

All existing methods for IP geolocation rely on the use of fixed hosts as
landmarks. In addition, in almost all cases, both landmarks and targets belong
to research facilities. The rationale for these choices may be found in the
difficulties concerning the setup of large-scale distributed experiments. In
this paper a novel method for active IP geolocation based on mobile devices,
enrolled according to crowdsourcing principles, has been presented.  For the
first time the use of smartphones is considered for the implementation of a
distributed IP geolocation platform, leveraging on the self-localization
ability of these devices. Since smartphones are crowdsourced from volunteers
who participate from largely different contexts, they belong to a wide range of
autonomous systems. Thus, the considered scenario is characterized by increased
heterogeneity with respect to previous research.  We believe that results
obtained in this scenario are more representative of real-world conditions. As
smartphones are connected to the Internet using wireless links, we devised
delay-distance models more suitable for this context than the ones available
from literature (calibrated for wired scenarios).  The model that takes into
account the access technology is able to provide better accuracy, but only when
the distance between the two endpoints is not too large. On the contrary, when
the distance between the two endpoints is significant, the benefits provided by
technology-calibrated models get almost nullified by the increased variability
of measurements. 

\section*{References}

\bibliographystyle{elsarticle-num}
\bibliography{mobgeo,locbiblio}

\end{document}